\documentclass{article}
\usepackage{xcolor}
\usepackage{bm,ulem}
\usepackage{amsmath}
\usepackage{graphicx} % if you want to include jpeg or pdf pictures
\usepackage{authblk}
\usepackage[colorlinks=true, linkcolor=blue, urlcolor=blue,
  citecolor=black]{hyperref}

\usepackage{amssymb}
\usepackage{lineno}
\usepackage{epstopdf, epsfig}%https://www.overleaf.com/project/65129455e0d1233c9af2bee9
\usepackage{appendix}

\newcommand{\beq}{\begin{equation}}
\newcommand{\eeq}{  \end{equation}}
\newcommand{\beqa}{\begin{eqnarray}}
\newcommand{\eeqa}{  \end{eqnarray}}
\usepackage{amsmath}

\newcommand{\grad}{{\boldsymbol \nabla}}
\usepackage{authblk}
\graphicspath{{../}}

\title{Cone-jet Stokes solutions in strong viscous flows: the vanishing flow rate limit}

\author[1,2,*]{Alfonso M. Ga{\~n\'a}n-Calvo}
\author[1]{Miguel A. Herrada}
\author[3]{Jens Eggers}

\affil[1]{Dept. Ing. Aerospacial y Mec{\'a}nica de Fluidos,Universidad de Sevilla.\\
Camino de los Descubrimientos s/n, 41092 Sevilla, Spain.}
\affil[2]{ENGREEN, Laboratory of Engineering for Energy and Environmental Sustainability, Universidad de Sevilla, 41092 Sevilla, Spain.}
\affil[3]{School of Mathematics, University of Bristol, Fry Building, Woodland Road, Bristol BS8 1UG, UK}

\begin{document}
\maketitle
*email: amgc@us.es

\begin{abstract}
Steady tip streaming in the vanishing flow rate limit has been evidenced both experimentally and numerically in the literature. However, local conical Stokes flow solutions supporting these results at vanishing small scales around the emitting tip have remained elusive. This work presents approximate local conical solutions in liquid-liquid flow focusing and tip streaming, in general, as the limit of a macroscopic vanishing issued flow rate. This provides mathematical foundations for the existence of an asymptotically vanishing scale at the tip of an intermediate conical flow geometry with angle $\alpha$. For a sufficiently small inner-to-outer liquid viscosity ratio $\lambda$, these solutions exhibit a universal power-law relationship between this ratio and the cone angle as $\alpha=k \lambda^{1/2}$, where the prefactor $k$, of the order of unity, depends on the geometric details of the macroscopic flow. This confirms the existing proposals that anticipate the use of flow focusing and tip streaming technologies for tight control of microscopic scales, down to those where diffuse liquid-liquid interfaces become manifested.
\end{abstract}

\section{Introduction}

Tight control of minute microscopic scales, even close to the molecular scale, can be considered the key to many of today's dominant technologies, ranging from pharmaceuticals, biochemistry or analytical chemistry to microelectronics. The fundamental step in realizing such control always involves the ability to {\sl focus} a stream of matter, be it a liquid, a gas or a photon beam, into a tightly controlled jet on a minute scale. When dealing with a liquid, focusing a differentiated stream of matter can be accomplished by several procedures involving a diversity of driving forces \cite{MG20}, which in almost all cases act against the surface tension of the liquid.

In general, the action of focusing is ultimately associated with a conical geometry, where mass flow or energy reach a singularity at its apex. In exceptional occasions, a purely conical self-similar solution of the balance equations is discovered. Indeed, \cite{B72} criticized the much earlier proposal by \cite{T34} where the latter suggested that pointed-apex bubbles in a straining flow could develop a conical tip if the phenomenon is observed with sufficient resolution. That possibility was experimentally entertained by \cite{RM61}. However, Buckmaster showed a fact already noted by \cite{T66b}: That a conical flow around a capillary hollow cone is not possible fundamentally because the mechanical balance of surface stresses (normal and tangential stresses should vanish on the bubble surface) is impossible at a perfect conical shape. \cite{E21} obtained the complete nonconical solution to the problem of the strained bubble tip shape with finite curvature apex in the absence of emission, and showed it could be matched to Taylor's theory as an outer solution.

\cite{T64} also discovered a conical, self-similar electrostatic solution (i.e. motionless) that served as the keystone of the so-called Taylor cone-jet electrospray mode, one of the major breakthroughs in analytical chemistry and biochemistry \cite{FMMWW89}. This procedure uses electrohydrodynamic forces to focus a liquid stream into a steady capillary jet whose diameter can be precisely determined on a scale ranging from millimeter to molecular \cite{Z17,CP89,F07,GLHRM18}. However, the singularity at the tip of the cone in Taylor's solution, which leads to perfectly balanced but unbounded local forces, prevents its strictly self-similar nature in practice, since the system eventually leads to the emission of matter from the tip of the cone. Nevertheless, Taylor's solution remains a solid keystone to the ultimate understanding of the physical phenomenon. In fact, the key to achieving a perfect balance of normal surface stresses is to counteract the surface tension with a normal electric stress. Taking advantage of this fact, other conical solutions have been reported assuming Stokes flow and the action of electrohydrodynamic forces \cite{RC94b,GM21}.

Inspired by electrosprays, \cite{G98a} discovered a way to focus a liquid stream by a converging gas flow in the form of a tiny capillary jet. The method, called {\sl Flow Focusing} (FF),  was later extended to produce the same focusing effect on a gas stream exerted by a converging liquid flow \cite{GG01}. The most popular configuration of FF was proposed by \cite{ABS03}, leading to a fundamental and long-lasting microfluidic paradigm. In reality, FF is part of a substantial class of fluid dynamics problems where a given volume of liquid is subject to stretching under a generalized extensional or focusing flow \cite{S94,Z04,CE06,SB06,GGRHF07,EC09,E21}. However, the analytical counterpart in FF or the steady tip streaming to the self-similar Taylor conical solution in electrospray remains unknown, despite a number of experimental and numerical evidences pointing to this possibility \cite{Z04,SB06,GGRHF07,Dong2018,CE20,RMEH24}. This solution would resolve a long-standing debate and the theoretical conundrum of whether simple mechanical means can effectively control flow scales down to the molecular level, leading to extreme mixing and emerging macroscopic properties such as those of ultra-fine emulsions (e.g. mayonnaise) despite the finite surface tension between the two phases involved.

We address a fundamental question in fluid physics arising from hydrodynamic focusing in the limit of a vanishing emitted flow rate: the existence of a local, self-similar conical flow structure that bridges two inherently distinct scales—a macroscopic domain influenced by non-conical boundary conditions, such as externally imposed extensional viscous flows, and a significantly smaller scale near the apex of the tapering meniscus. To resolve this paradox, two complementary analytical approaches are pursued. The first employs general solutions of the Stokes equations in spherical coordinates, supplemented by numerical analysis to accurately capture the non-conical region that connects the self-similar cone to the emerging jet. In contrast, the second approach utilizes slender-body lubrication theory, analytically capturing the complete cone-jet structure autonomously, provided the cone angle is sufficiently small. Remarkably, the lubrication theory yields a universal self-similar flow structure governed by a single dimensionless parameter directly linked to the extensional strength of the outer focusing fluid. Both solutions exhibit the same scaling dependency of the cone angle on the square root of the viscosity ratio between the inner and outer fluids. Ultimately, these two methods yield results in nearly perfect mutual agreement, with their correspondence becoming exact as the cone angle approaches zero. By analytically and numerically resolving this conundrum, we establish the feasibility of a locally conical, infinitely thinning flow geometry with virtually negligible emission, serving as an ideal intermediate asymptotic structure for forming perfectly cylindrical jets at scales approaching the ultimate continuum limit.

%Here we propose that such a universal quasiconical solution with virtually zero emission exists and can be approached both analytically and numerically. In addition, its stability can also be studied.

The first method is structured in two steps:

1- We first introduce an analytical conical flow solution of the Stokes equations as a local solution of flow focusing or tip-streaming in the limit of a zero emitted flow. This solution is a modification of Buckmaster's solution \cite{B72} to include the internal recirculating flow of the liquid inside the cone, which allows an exact balance of viscous stresses with surface tension at the cone surface with semi-angle $\alpha$ for a wide range of viscosity ratios between the two fluids and cone angles. Interestingly, this solution yields a {\sl constant} velocity (independent of the spherical radial coordinate) at the cone surface. However, it requires the axis to be avoided in the external region due to the logarithmic singularity appearing there. Physically, the local flow around the axis emanating from the cone is akin to an artificial infinitely thin "drawing line'' in the external flow moving the liquid from the tip of the cone to keep the flow running in the axial direction. The appearance of the logarithmic singularity at the axis in the flow outside the cone shows that a line of point forces or stokeslets at the axis must exist to keep the outer flow going in the form of an infinite conical configuration with a constant surface speed. However, this solution establishes the basis for a subsequent valid approximation: It is the extra degree of freedom at the axis that actually allows the existence of a conical solution, as illustrated next.

2- The above conical solution is superimposed with another inner solution having a content flux to account for the emitted flow rate and is approximately matched, at large distances from the cone-jet transition, with a solution considering an asymptotically cylindrical infinite jet with a diameter proportional to the local cone-jet transition scale.
%This determines the asymptotic cone angle as a function of the viscosities ratio.
The complete solution at the tip of the cone naturally entails the breakdown of the conical self-similar solution at the local scale of the cone-jet transition, which is commensurable with the diameter of the emitted jet. At this scale, the conical shape tapers into a universal funnel shape that eventually turns into a perfectly cylindrical jet. This transition flow region is finally solved numerically for a given viscosity ratio and cone angle by Herrada's method \cite{HM16a} using the proposed analytical solution as the asymptotic boundary condition.

The second method expands about the local flow strength at the axis of an arbitrary external flow governed by the axisymmetric Stokes equations. Using the slender-body approximation, a general formulation based on lubrication theory is developed, which is reduced to a similarity solution describing the cone-jet structure. This similarity solution is characterized by a single dimensionless parameter that represents the scaled external flow velocity at the point on the axis where the conical region transitions into a cylindrical thread. This governing parameter, defined as proportional to the square root of the viscosity ratio between the inner and outer fluids, exhibits a threshold below which stable conical solutions exist. This result is consistent with the predictions from the first approach: Remarkably, this threshold corresponds to the cone angle that maximizes the interface velocity for the analytically exact solution, in the limit of vanishing viscosity ratio. Beyond this threshold, the solutions become non-conical (cusp-like), and their analysis lies outside the scope of the present paper.

\section{Problem formulation}
\label{formulation}

The problem we address assumes the presence of a stretching or converging flow in an incompressible and viscous fluid. This flow can be produced by any procedure or geometry of the contours, for example, a liquid stream forced through an orifice such as the FF configuration, or an extensional axisymmetric flow. Within this flow pattern, another fluid immiscible with the first and forming a capillary meniscus (for example, a droplet \cite{SL89b}) is subjected to stretching by the viscous action of the first fluid to the point where, at the apex of the stretched meniscus, a capillary jet of characteristic diameter much smaller than any other scale existing in the system is emitted.

In this study, we will address the physical question of whether a self-similar flow pattern and a conical meniscus shape is possible as the intermediate geometry of the tip of the meniscus in question. If such a flow solution exists, it is the perfect candidate to serve as the source of a perfectly cylindrical jet with a diameter that can be as small as the scale of the continuum hypothesis \cite{GGRHF07}.

%since the only geometry that admits a flow configuration with a vanishing-scale tip located at a fixed point is a cone.

\subsection{Physical scales and intermediate range}
\label{scales}

%First, we are not addressing the stability of the global flow at the large spatiotemporal scales, much larger than those of the problem at hand. We also discard local solutions of the mathematical ``cusp" type, since these are not compatible with self-similarity. In other words,

\begin{figure}
  \centerline{\includegraphics[width=0.99\textwidth]{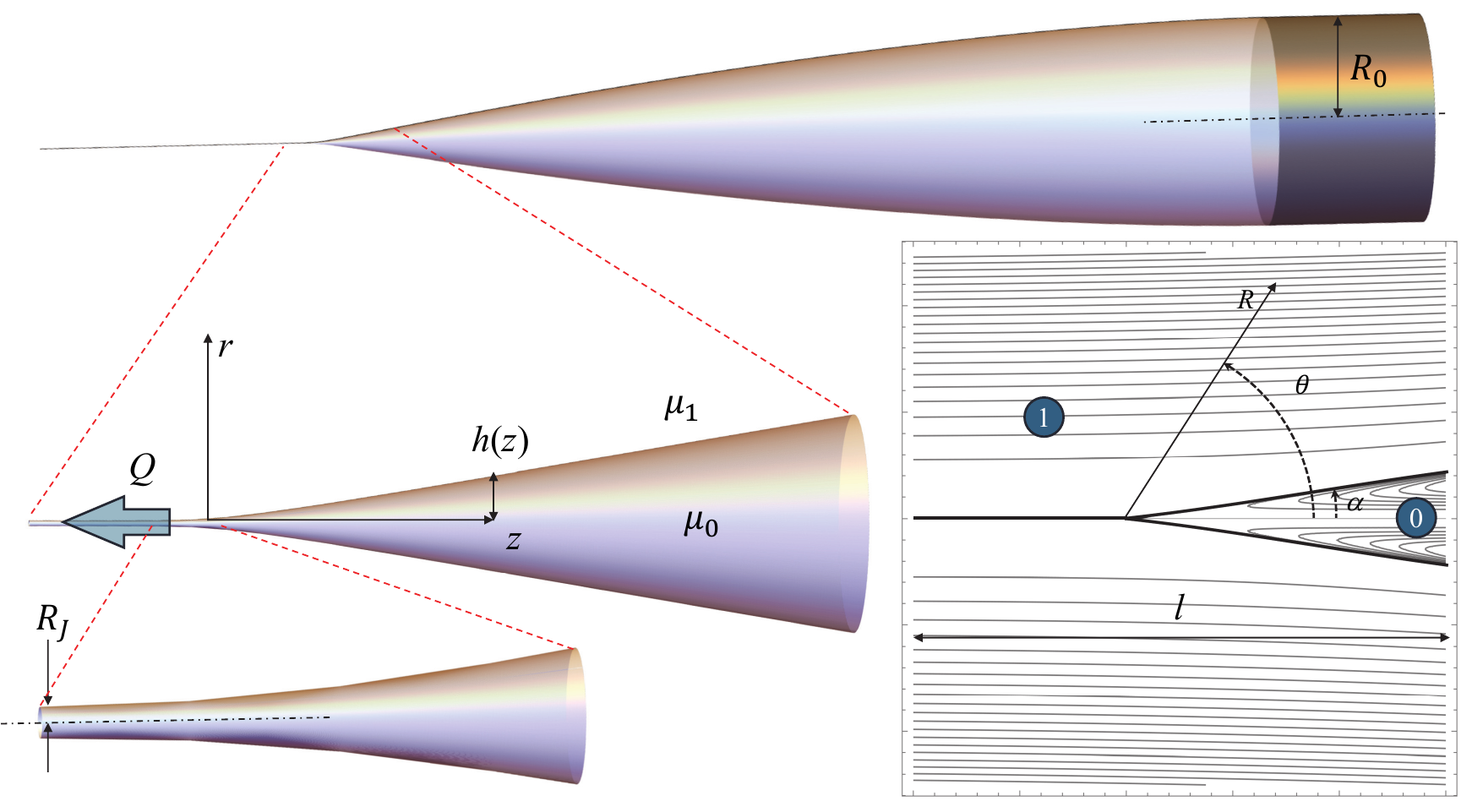}}% Images in 100% size
  \caption{The global flow focusing geometry, the intermediate scale where a conical geometry emerges, and the idealized local conical flow  considered in this work. The local spherical coordinates $R$ and $\theta$ are indicated, as well as domains 0 (meniscus tip) and 1 (focusing stream). Also, the cylindrical coordinates $\{r,z\}$ and the meniscus profile radius $h(z)$  used in the slender body theory are indicated. $\mu_0$ and $\mu_1$ are the viscosities of the inner and outer incompressible fluids, respectively. The macroscopic scale $R_0$ is imposed by any external boundary condition (here, a feeding tube), while the intermediate scale $l$ denotes any length scale below $R_0$ around the tip where a local conical meniscus can be observed. $Q$ is the ejected flow rate of the inner fluid. $R_J$ is the jet radius once the quasi-cylindrical geometry is developed.}
\label{f1}
\end{figure}

We seek the existence of an intermediate scale range in which a solution dominated by viscosity (see Figure \ref{f1}) can realistically be represented by a stable conical flow. That geometry would exist between the large global macroscopic scale and the very small scale in the region that smoothly links the conical shape with the formally infinitesimal emitted capillary jet. In that intermediate scale range, if the solution sought existed, the flow would be self-similar. Globally stable solutions would be supported in the parametric region where this flow would be locally stable.

\subsubsection{Is there a conical intermediate region?: Local, intermediate, and viscous scales}
\label{sub:local}

The existence of a conical local and universal solution can be demonstrated if we use a numerical solution with realistic boundary conditions at the macroscopic scale and find such an intermediate region of invariance. In the recent work of \cite{RMEH24}, the authors use a configuration with an external extensional flow that does not support asymptotic conical solutions for large scales. Despite this, an intermediate but very large scale compared to the size of the cone-jet region can always be found when the issued flow rate vanishes.

Thus, we seek the conical flow solution of an incompressible viscous fluid stream with viscosity $\mu_0$ focused by a second immiscible and axially symmetric viscous co-flow with viscosity $\mu_1$, in the limit of a vanishing ejected flow rate $Q$ in units of a macroscopic length scale. In this limit, we hypothesize the existence of an intermediate conical shape of characteristic size $l$ near the tip of the focused meniscus, from which a steady ejection takes place as a very thin capillary jet.

If the driving outer flow is locally inertia-less in the vicinity of the cone-jet transition, the local scale of the issued stream (capillary jet) can be described by a characteristic length $l_0$, exclusively determined by the three local dominant parameters, namely the viscosity of the outer driving fluid $\mu_1$, the ejected volume flow rate $Q$ and the surface tension $\gamma$ between the outer and inner fluids, as $l_0=\left(\frac{\mu_1 Q}{\gamma}\right)^{1/2}$. Thus, at this scale the problem will be written in units of length, mass and time as $l_0$, $m_0=\frac{\mu_1^3 Q}{\gamma^2}$ and $t_0=\left(\frac{\mu_1^3 Q}{\gamma^3}\right)^{1/2}$, respectively. These length, mass and time scales would vanish when the flow rate vanishes as measured with units of the conical intermediate scale $l \gg l_0$. In contrast, interestingly, the scales of pressure
$p_0 = \left(\frac{\gamma^3}{\mu_1 Q}\right)^{1/2}$
and density
$\rho_0=\left(\frac{\mu_1^3}{\gamma Q}\right)^{1/2}$
diverge as $Q^{-1/2}$, since the pressure should balance the surface tension, i.e. $\left(\gamma/l_0\right)p_0^{-1}=1$, and the density ratio $\rho/\rho_0 \ll 1$ indicates an inertia-less flow. In contrast, in units of $l_0$, $m_0$ and $t_0$, the values of the surface tension and flow rate are $\gamma = 1$ and $Q=1$, respectively, and the values of the inner and outer viscosities are $\lambda_0=\mu_0/\mu_1$ and $\lambda_1=1$. From now on, whenever $\lambda$ appears without sub-index, it means $\lambda_0$.

To ensure that viscous focusing forces dominate over surface tension, the characteristic length scale $l_0$ must satisfy $l_0 \ll l_\mu = \frac{\mu_1^2}{\rho_1 \gamma}$, where $\rho_1$ is the density of the outer focusing fluid. Introducing the characteristic flow rate $Q_\mu = \left(\frac{\mu_1^3}{\rho_1^2 \gamma}\right)^{1/2}$, this condition implies that the nondimensional flow rate must satisfy $Q/Q_\mu \equiv \left(\frac{\rho_1^2 Q \gamma}{\mu_1^3}\right)^{1/2} \ll 1$. Additionally, for a viscous-dominated conical flow pattern to emerge, one must consider an intermediate length scale $l$, loosely defined as any scale fulfilling $l_0 \ll l \ll l_\mu$.
Therefore, using the intermediate scale $l$ and the units of $\mu$ and $\gamma$, introducing the reference flow rate $Q_l=\gamma l^2/\mu $ one should also have $Q \ll Q_l \ll Q_\mu$, consistently with $l_0 \ll l \ll l_\mu$.

% Units

From a problem-solving point of view, the practical use of these length scales and their corresponding units can be reduced to considering the value of the flow rate as follows:

1- At the intermediate scale $l$ and time $t_l=\mu l/\gamma$ units, the flow rate can be given by the value $q\equiv Q/Q_l$. From this scale, in the limit $q \rightarrow 0$ there would be no visible jet and the outer domain flow (1) must comply with regularity conditions at the axis, which will be subsequently given in detail.

2- At the local scale, using $l_0=\left(\mu_1 Q/\gamma\right)^{1/2}$ and $t_0$ as the unit length and time, respectively, one necessarily has $Q=1$.

In summary, as long as one had $l\gg l_0$, one would have $q\ll 1$, but if $l$ approaches $l_0$, in the limit $l=l_0$ one has $Q=q=1$ (again, in units of $l_0$ and $t_0$).

Next, we discuss the equations and boundary conditions of the problem in spherical polar coordinates.

\subsection{Conical Stokes flow and boundary conditions}

% Stream function and velocity field
In the inertia-less limit, we use the stream function $\Psi_j(r,\theta)$ for the Stokes flow in the focused ($j=0$) and focusing ($j=1$) domains (these initial domains can be extended subsequently). This function is determined by the well-known differential equation \cite{Happel1973,Liu1978}:
\begin{equation}
\label{e1}
E^2(E^2 \Psi_j)=0.
\end{equation}
We choose spherical coordinates $\{R,\theta\}$, where the operator $E^2$ obeys the expression
\begin{equation}
E^2\equiv \frac{\partial^2}{\partial R^2}+\frac{\sin\theta}{R^2}\frac{\partial}{\partial \theta}\left(\frac{1}{\sin\theta}\frac{\partial}{\partial\theta}\right).
\end{equation}
The velocity field $\mathbf{u}=\left\{ u_R,u_\theta\right\}=\{\frac{1}{R^2\sin\theta}\frac{\partial\Psi}{\partial\theta} ,-\frac{1}{R\sin\theta}\frac{\partial\Psi}{\partial R}\}$, the components of the stress tensor $\boldsymbol{\tau}$,  $\tau_{R,R}=2\lambda_j \frac{\partial u_R}{\partial R}$, $\tau_{R,\theta}=\lambda_j \left( R \frac{\partial (u_\theta/R)}{\partial R} + \frac{1}{R} \frac{\partial u_R}{\partial\theta} \right)$, and $\tau_{\theta,\theta} = 2 \lambda\left( \frac{\partial u_\theta}{\partial \theta} +u_R\right)R^{-1}$ and the pressure $p$ at each domain $j$ satisfy the Stokes equation $\nabla p_j= \nabla \cdot \boldsymbol{\tau}$.

% Conditions at the axis

Singularities at the axes $\theta = 0$ and $\theta = \pi/2$ should be avoided in both the inner and outer domains of the cone ($j=0$ and 1, respectively), that is,

\begin{equation}
    u_{\theta}^{(0)}(R,\theta=0)=0,\, \frac{\partial u_{R}^{(0)}}{\partial \theta}(R,\theta=0)=0, \, u_{\theta}^{(1)}(R,\theta=\pi)=0,\, \frac{\partial u_{R}^{(1)}}{\partial \theta}(R,\theta=\pi)=0
    \label{axis}
\end{equation}

% Boundary conditions and flow rate

At the meniscus surface $\theta=\theta_s(R)$, where the normal and tangential unit vectors are expressed as
\[\mathbf{n}=\{-R\, \theta'_s(R),1\}\left(1+R^2\theta'_s(R)^2\right)^{-1/2}\]
 and
 \[ \mathbf{t}=\{1,R\, \theta'_s(R)\}\left(1+R^2\theta'_s(R)^2\right)^{-1/2},\]
respectively -- the primes indicate derivatives with respect to the variable indicated ($R$ in this case)-- the normal and tangential stress balance read:
\begin{equation}
p^{(1)}-p^{(0)}+\mathbf{n}\cdot (\left(\boldsymbol{\tau}^{(0)}-\boldsymbol{\tau}^{(1)}\right)\cdot \mathbf{n}) + \kappa(R)=0,
\label{normal}
\end{equation}
and
\begin{equation}
\mathbf{t}\cdot\left( \left(\boldsymbol{\tau}^{(0)}-\boldsymbol{\tau}^{(1)}\right)\cdot \mathbf{n}\right)=0
\label{tangential}
\end{equation}
where the curvature $\kappa(R)$ in spherical coordinates is given in general by:
\begin{equation}
    \kappa(R)= \frac{\cot (\theta_s (R))-R \left(r \theta''_s(R)+\theta'_s(R) \left(R\, \theta'_s(R) \left(2 R\, \theta'_s(R)-\cot (\theta_s (R))\right)+3\right)\right)}{R \left(R^2 \theta'_s(R)^2+1\right)^{3/2}},
    \label{curvature}
\end{equation}
which is reduced to $\kappa=\frac{\cot (\theta_s (R))}{R}$ for a conical shape.

The continuity of tangential velocities requires
\begin{equation}
\left(\mathbf{u}^{(1)}-\mathbf{u}^{(0)}\right) \cdot \mathbf{t}=0.
\label{ut}
\end{equation}

For a steady interface $F(R,\theta)\equiv \theta_s(R)-\theta=0$, the general requirement $\frac{\partial F}{\partial t} + \mathbf{u}\cdot \nabla F=0$ demands:
\begin{equation}
\mathbf{u}^{(0)}\cdot \mathbf{n} = \mathbf{u}^{(1)}\cdot \mathbf{n}=0.
\label{un}
\end{equation}

Recalling that we are using a generic length scale $l$, and $\mu_1$ and $\gamma$ as the units of viscosity and surface tension, for any value of $R$, the net flow rate is given by
\begin{equation}
q=\int_0^{\theta_s(R)} 2 \pi R^2 u_R^{(0)} \sin (\theta ) \, d\theta
\label{Q}
\end{equation}
Alternatively, as anticipated, using the local scale $l_0=\left(\mu_1 Q/\gamma\right)^{1/2}$, the flow rate is set to $q=-1$.

\section{Proposed solutions}\label{solution}

\subsection{A first analytical conical exact solution}
\label{firstsol}

Analytical solutions of equation (\ref{e1}) in separate variables have been known long ago. The solution part depending on the angular coordinate $\theta$ can be expressed as combinations of four independent solutions to the fourth-order differential equation derived from (\ref{e1}), which we express for reference in terms of the associated Legendre functions as \cite{B72,Happel1973,Liu1978}:
\begin{eqnarray}
\Psi_j = R^{3/2+\beta} f^{(j)}(x) \equiv R^{3/2+\beta} (1-x^2)^{1/2} \left( A^{(j)}_1(\beta) P_{\beta+1/2}^1(x) + \right. \nonumber \\ A^{(j)}_2(\beta) P_{\beta-3/2}^1(x) +
\left. A^{(j)}_3(\beta) Q_{\beta+1/2}^1(x) + A^{(j)}_4(\beta) Q_{\beta-3/2}^1(x)\right),
\label{gsol}
\end{eqnarray}
where $x=\cos(\theta) $, and $P_\nu^1(x)$ and $Q_\nu^1(x)$ are the associated Legendre functions.

The assumed existence of the intermediate region does not automatically suggests that we can set a perfectly conical flow at infinity, except in the case of the limit $q\rightarrow 0$. In this case, there are {\sl two} independent parameters of the problem: the viscosity ratio $\lambda$ and the cone angle $\alpha$.

\begin{figure}
  \centerline{\includegraphics[width=0.50\textwidth]{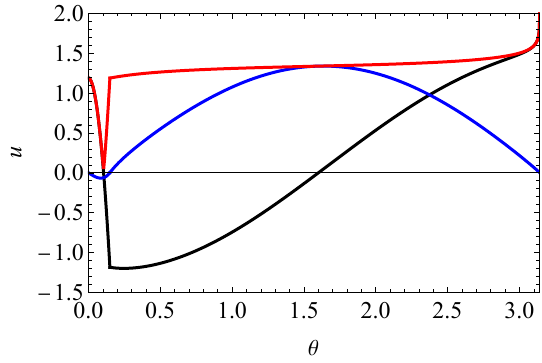}\includegraphics[width=0.50\textwidth]{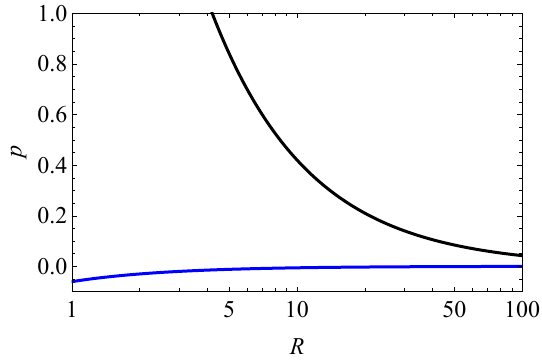}}% Images in 100% size
  \caption{(Left): The fluid velocities $u_R^{(j)}(\theta)$ (black curves) and $u_\theta^{(j)}(\theta)$ (blue curves) of the exact analytical solution (\ref{psol})-(\ref{As}), plotted as functions of $\theta$. The red curve represents the modulus of the velocity. Note the singular behavior as $\theta \rightarrow \pi$ (the region that would be occupied by the jet). (Right): The pressure distributions $P^{(0)}(R)$ (black) and $P^{(1)}(R)$ (blue). Here, $\alpha=0.15$, $\lambda=0.01$, and $q=0$.}
\label{f2}
\end{figure}

Thus, reducing the problem to a perfectly conical self-similar flow with a meniscus of semi-angle $\theta_s(R)=\alpha$, solution to equations (\ref{gsol})-(\ref{Q}) can be written as a superposition of two similarity solutions  \cite{B72,Happel1973}:
\begin{eqnarray}
\Psi_j = R^2 \left(G_{j,1}+G_{j,2} \cos (\theta ) + G_{j,3}\cos ^2(\theta ) + G_{j,4}\sin ^2(\theta ) \tanh ^{-1}(\cos (\theta ))\right)
\nonumber \\
+ A_{j,2}\cos ^3(\theta )+ A_{j,1}\cos (\theta )+ A_{j,3} \cos (\theta ) \left(\cos (\theta )-\sin ^2(\theta ) \log \left(\tan \left(\frac{\theta }{2}\right)\right)\right),
\label{psol}
\end{eqnarray}
The terms proportional to $R^2$ ($\beta=1/2$,  coefficients $G_{j,i}$: the ``stress solution") satisfy the stress conditions and those independent of $R$ ($\beta=-3/2$, coefficients $A_{j,i}$: the ``flux solution") provide the flow rate. In this first solution, the first index of each set labels regions (0) and (1), the second the amplitude of the linearly independent solutions; the fourth term in the solution independent of $R$ is a constant and then can be dropped. As a result, there are only three coefficients $\left.A_{j,i}\right|{1,2,3}$. The boundary conditions lead to:
\begin{eqnarray}
G_{0,1} & = & -\frac{\cos ^2\left(\alpha\right) \cot \left(\frac{\alpha}{2}\right)}{4 \left(\left(\lambda-1\right) \cos \left(\alpha\right)+\lambda+1\right)}, \nonumber \\
G_{0,2} & = & \frac{\sin \left(\alpha\right) \cot ^2\left(\frac{\alpha}{2}\right)}{4 \left(\left(\lambda+1\right) \sec \left(\alpha\right)+\lambda-1\right)}, \nonumber \\
G_{0,3} & = & -\left(G_{0,1}+G_{0,2}\right),\,\,
G_{0,4} = 0,\nonumber \\
G_{1,1} & = & -\frac{\lambda A -B}{8 \left(\left(\lambda-1\right) \cos \left(\alpha\right)+\lambda+1\right)}, \nonumber \\
G_{1,2} & = & -\frac{\sin ^2\left(\frac{\alpha}{2}\right) \cos \left(\alpha\right) \tan \left(\frac{\alpha}{2}\right) \left(\left(\lambda-1\right) \cos \left(\alpha\right)+\lambda\right)}{2 \left(\left(\lambda-1\right) \cos \left(\alpha\right)+\lambda+1\right)}, \nonumber \\
G_{1,4} & = & \frac{1}{8} \sin \left(2 \alpha\right), \,\,
G_{1,3} = -\left(G_{1,1}-G_{1,2}\right),
\label{Gs}
\end{eqnarray}
being
\begin{eqnarray}
A & = & \sin \left(2 \alpha\right) \left(\cos \left(\alpha\right)+\left(\cos \left(\alpha\right)+1\right) \log \left(\tan \left(\frac{\alpha}{2}\right)\right)\right) \nonumber \\
    B & = & \left(\sin \left(\alpha\right)-\tan \left(\frac{\alpha}{2}\right)\right) \left(\cos \left(2 \alpha\right)-2 \sin ^2\left(\alpha\right) \log \left(\tan \left(\frac{\alpha}{2}\right)\right)+1\right).
\end{eqnarray}
In the general case with an ejected flow rate $q$, one has
\begin{equation}
A_{j,1 } = -\frac{3 \lambda_j \cos ^2\left(\alpha\right) \csc ^4\left(\frac{\alpha}{2}\right)}{16 \pi  \cos \left(\alpha\right)+8 \pi }q, \quad
A_{j,2}  =  \frac{\lambda_j \csc ^4\left(\frac{\alpha}{2}\right)}{16 \pi  \cos \left(\alpha\right)+8 \pi }q,\quad
A_{j,3} = 0.
\label{As}
\end{equation}
for $j=0,1$. The coefficient $A_{j,3}$ is zero since the solution of the form
\begin{equation}
A_{j,3} \cos (\theta ) \left(\cos (\theta )-\sin ^2(\theta ) \log \left(\tan \left(\frac{\theta }{2}\right)\right)\right),
\label{A3}
\end{equation}
which gives a strong singularity on the axis, should be excluded in both domains. On the other hand, when $R \gg 1$, both $u_R$ and $u_{\theta}$ are independent of $R$, as required, but they exhibit a logarithmic singularity as $\theta \rightarrow \pi$, i.e. the region occupied by the jet. An illustrating plot of them as functions of $\theta$ for $R \rightarrow \infty$ is given in figure \ref{f2}.

This flow pattern raises the inner liquid pressure towards the apex: When the inner-to-outer viscosity ratio is small, the momentum diffusion from the outer converging flow is so strong that it compresses the inner flow at the tip. In effect, for $R\gg 1$, the corresponding expressions for the pressure in both domains are given by:
\begin{eqnarray}
p_j & = & \lambda_j \frac{2 \left(G_{j,2}-G_{j,4}\right)}{R}.
\label{pressure}
\end{eqnarray}
This expression is positive for $j=0$ and negative for $j=1$, and becomes unbounded for $R\rightarrow 0$. Note that, in contrast with the fluid velocities, the pressure is independent of $\theta$. Figure \ref{f2}(right) illustrates the pressure distributions in both domains, which are inversely proportional to $R$ to balance the surface tension term.

For the given small viscosity ratio $\lambda=0.01$ used in figure \ref{f2}, the pressure distribution is due to the intense diffusion of momentum into the inner flow from the outer domain, so that while the outer flow loses pressure only modestly, the inner flow gains it substantially. Their difference is balanced by the increase in surface tension and normal viscous forces as $R$ decreases. The increasing overpressure towards the apex in the inner domain projects the on-axis stream in the opposite direction to that of the interface towards the apex.

\subsubsection{The interfacial velocity}
\label{intervel}

From the solution (\ref{psol})-(\ref{As}), the velocity at the interface reads:
\begin{equation}
    u_R(\alpha)=-\frac{\sin \left(2 \alpha\right)}{8 \left(\left(\lambda-1\right) \cos \left(\alpha\right)+\lambda+1\right)}
  \label{US}
\end{equation}

\begin{figure}
  \centerline{\includegraphics[width=0.50\textwidth]{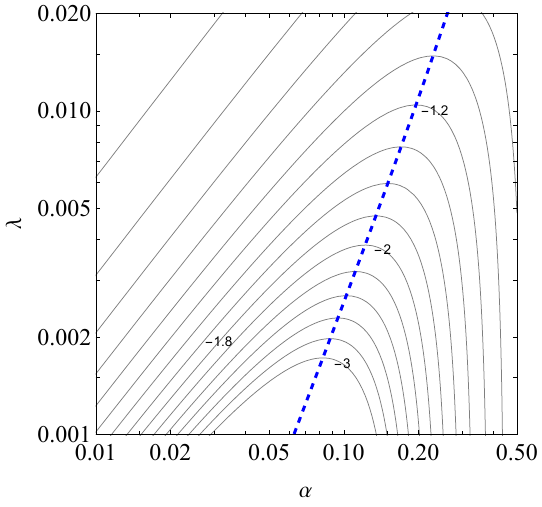}}
  \caption{The velocity of the interface (in the direction of the apex) according to (\ref{US}), as a function of $\alpha$ and $\lambda$. Iso-contours represent constant velocity values. The blue dashed line is the value of $\alpha$ that maximizes the absolute value of the interfacial velocity for a given viscosity ratio $\lambda$. Above this maximum absolute velocity for a given $\lambda$, there is no solution, while below it, each velocity gives two possible solutions and cone angles. In section \ref{Jens} we show that the solutions to the right of the blue curve should not be considered.
  %Red line is the fitting $\lambda=\alpha^2(1-1.077\alpha)$.
  }
\label{f4}
\end{figure}

This velocity can be represented as a function of the cone angle $\alpha$ and the viscosity ratio $\lambda$. It is given in figure \ref{f4}. Note that the modulus of the outer velocity is nearly constant everywhere in the outer domain, except at the axis.

For a given driving velocity and a given viscosity ratio $\lambda$, Figure \ref{f4} shows that the cone angle may have two possible values, or no solution.
%Therefore, the exact analytic solution suggests that both $\lambda$ and $\alpha$ are free parameters of the problem. However, $\alpha$ is not completely free in this problem. In first place,
There is one value of $\alpha$ that maximizes the absolute value of the surface velocity for a given $\lambda$, represented as a blue dashed line: It is the cone angle for which the transfer of momentum to the inner fluid is maximized. This maximizing $\alpha_m$ value is given by the expression:
\begin{equation}
\lambda=\frac{2 \tan ^2\left(\frac{\alpha_m}{2}\right) \left(\sin ^2\left(\alpha_m\right)+\cos \left(\alpha_m\right)\right)}{2 \cos \left(\alpha_m\right)+\cos \left(2 \alpha_m\right)-1}.
\label{alphamax}
\end{equation}
For $\lambda\ll 1$, one has that (\ref{alphamax}) can be approximated as:
\begin{equation}
\lambda = \frac{\alpha_m^2}{4}\left(1+\frac{13}{6}\alpha_m^2\right)+O(\lambda^6) \Longrightarrow \alpha_m= 2 \lambda^{1/2}%\left(1-\frac{13}{3}\lambda^2\right)
+O(\lambda^{3/2}),
\label{lambda_scale}
\end{equation}
which is exactly the same relationship $\alpha_m = 2\lambda^{1/2}$ for the critical solution (maximum strength of the external flow) found in section \ref{Jens} using slender body theory.

However, given the logarithmic singularity of the axial velocity on the outer axis ($\theta = \pi$), the analytic solution presented in this section (see Figure \ref{f2}) does not completely solve our problem. Nevertheless, the existence of this solution suggests that such a conical flow configuration would be a good candidate as a first approach, modified as necessary to satisfy the presence of an inertialess thin jet in the vicinity of $\theta = \pi$ for small $\lambda$ values. The modification could be accomplished by forcing a regularization around the jet of the exact solution (\ref{psol})-(\ref{As}) presented. This possibility will be explored next.

\subsection{Cone-jet solution: an optimal approximate analytical solution with a jet domain}

\begin{figure}
  \centerline{\includegraphics[width=0.90\textwidth]{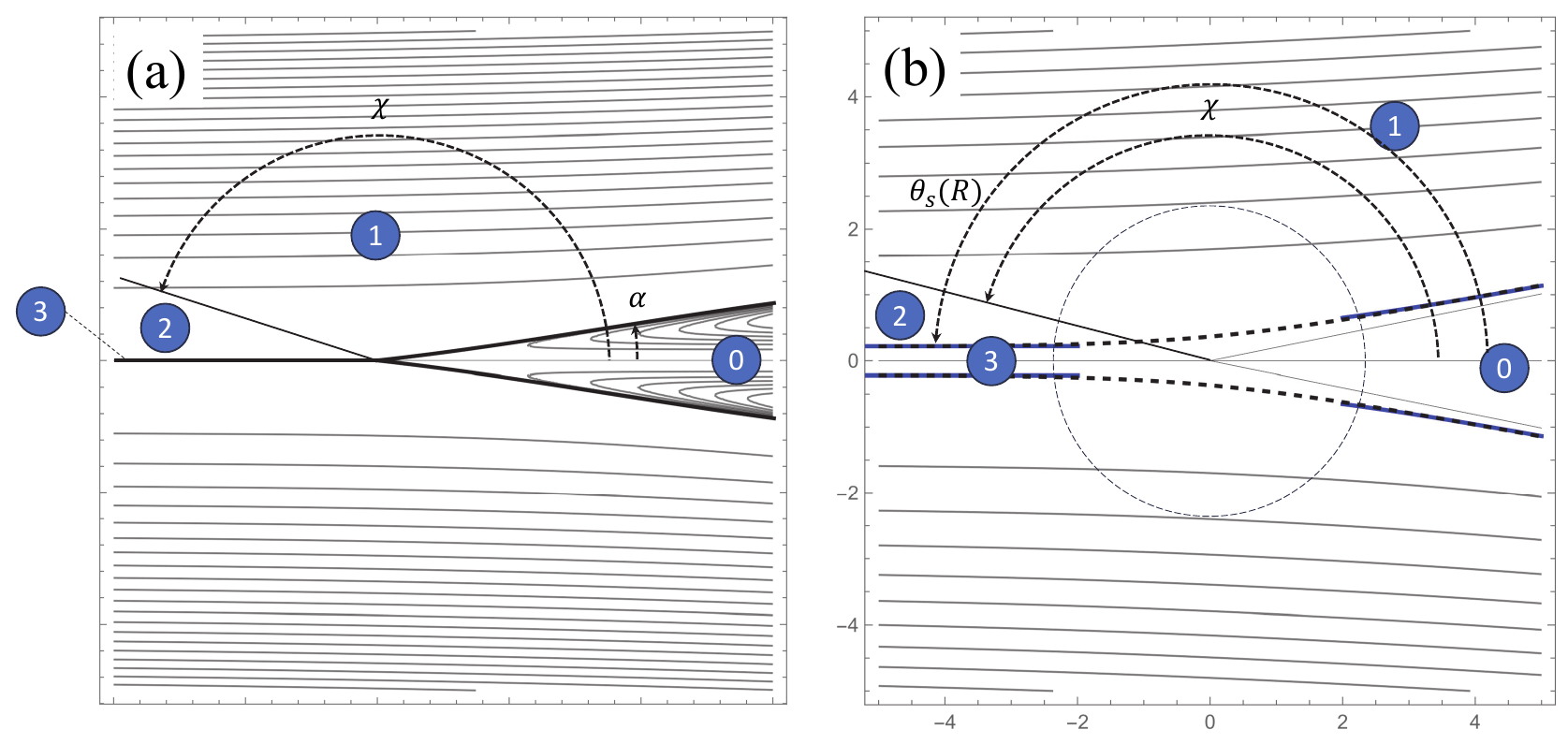}}% Images in 100% size
  \caption{The four regions considered: 0 (inner cone), 1 (outer cone), 2 (outer jet region) and 3 (inner jet). The angle $\chi$ separating regions 1 and 2 is where the solutions should match. According to the procedure described, this angle is a free parameter that determines the values of the cone angle $\alpha$ and the jet radius $R_J$ for a given viscosity ratio $\lambda$. It is related to the free parameter $\overline{\rm Ca}$ of the second solution procedure described in Section \ref{Jens}. (a) General schematics for an arbitrary intermediate scale $l$. (b) The analytical solution at the local scale $l_o$ (blue lines). The ultimate cone-jet transition is numerically resolved (dashed line).}
\label{regions}
\end{figure}

To tackle regularization, we divide the space into four domains (see Figure \ref{regions}):

1- The inner cone (domain (0)),

2- The outer flow to the cone (domain (1), now limited between the angles $\alpha$ and $\chi$),

3- The outer jet domain (2) (or the outer flow in the jet region, between the angles $\chi$ and $\pi-R_J/R$, for $R_J \ll R$), and

4- The inner jet domain (3).

In this part, we assume a non-zero flux ($q\neq 0$). Thus, a jet will be present and the self-similarity of the problem is lost in favor of a local solution of the cone-jet flow geometry with a finite scale of the jet radius. By choosing $q=1$, we implicitly use $l_0$ as the unit of length.

The analytical approach at the cone to solve domains (0) and (1), leading to solution (\ref{psol})-(\ref{As}) by setting the boundary conditions at the cone surface, is followed in the case of domains (2) and (3), which correspond to the jet problem. Once both the cone and jet problems are independently solved, the remaining unknowns to close the global problem can be solved by matching domains (1) and (2).

Focusing now on the jet surface, given by $\theta_s(R)=\pi - R_J/R$ in the limit of small $R_J$, Similarly to the cone domain we seek for solutions of the form:
\begin{equation}
    \psi_j=\cos ^3(\theta ) A_{j,2}+\cos (\theta ) A_{j,1}+R^2 \left(\cos ^2(\theta ) G_{j,3}+\cos (\theta ) G_{j,2}+G_{j,1}\right)
\end{equation}
with $j=2,3$, which complies with regularity at the axis. The singular part of the solution is dropped, i.e. $\left.G_{j,4}\right|_{j=2,3}=0$. The conditions on the jet surface are identical to (\ref{normal})-(\ref{un}), except for the value of $\theta_s(R)$. To match the flow from the cone side, i.e. condition (\ref{Q}), the flow rate condition in the jet domain for any radial position $R$ is now:
\begin{equation}
    q=\int_{\pi-R_J/R}^{\pi}2 \pi R\, \sin (\theta) \,u_R^{(3)}\, R\,d\theta = 1
    \label{Qj}
\end{equation}
since the flow is now positive in the $R$ direction.

As stated previously, the values of $A_{2,1}$ and $A_{2,2}$ can be identical to those obtained in the outside region of the cone, $A_{1,1}$ and $A_{1,2}$. This also guarantees that the stress component $\tau_{RR}$, given by
\begin{equation}
\tau_{RR}=\frac{4 \left(\cos (\theta ) \left(3 A_{1,2} \cos (\theta )-R^2 G_{1,4}\right)+A_{1,1}\right)}{R^3}
\end{equation}
is identical at both the cone and the jet regions, since no terms depending on $G_{j,i}$ appear in this expression. Moreover, to avoid tangential stresses $\tau_{R\theta}$ on the jet surface as $R \rightarrow \infty$ (i.e., the jet velocity profile should be flat), one should have $G_{j,4}= 0$ in the jet region. In contrast, $G_{j,4}\neq 0$ in cone region (1) (see expressions (\ref{Gs})), which implies a logarithmic singularity when $\theta \rightarrow \pi$ (or $R\rightarrow \infty$). Therefore, $G_{2,i}|_{i=1,2,3}$ cannot be identical to  $G_{1,i}|_{i=1,2,3}$, and a perfect match is impossible. However, the singularity on the axis diminishes with the cone angle since $G_{1,4}$ vanishes as $\alpha/4 \rightarrow 0$, while the other coefficients remain of order unity or even larger.
%This harbors the possibility of a sufficiently accurate solution for $\alpha \ll 1$.

In the following, a way to obtain a global solution by an optimal approximate analytical match between the two solutions of (\ref{gsol}) at the intermediate angle $\chi$ together with the solution on the inner side of the jet is presented.

Given a viscosity ratio $\lambda$, a set of 11 unknowns to solve the problem is given by the following:

1- Eight coefficients, namely $G_{j,i}|_{j=2,3; i=1,2,3}$ and $A_{3,i}|_{i=1,2}$,

2- The jet diameter $R_J$,

3- The cone angle $\alpha$, and

4- The angle $\chi$ where the errors between the solutions at domains (1) and (2) are minimal (approximate matching), or zero.

\subsubsection{Detailed procedure to calculate the remaining coefficients $G_{j,i}$ and $A_{3,i}$}

Recalling that in units of $l_0$ the variable $R$ is $R\gg 1$ in the intermediate conical region, the 11 equations necessary to resolve the 11 unknowns are obtained as follows.

1- After applying the boundary conditions on the jet axis (\ref{axis}) and neglecting the resulting terms smaller than $R^{-3}$ (that is, neglecting $R^{-4}$ and beyond), the condition of zero normal velocities at the jet surface (2 conditions) give 4 equations for terms of the order $R$ and $R^{-1}$.

2- Moreover, the continuity of tangential velocities (1 cond.) give 2 eqns. for terms as $R^0$ and $R^{-2}$.

3- Finally, the normal and tangential stresses at the jet surface (2 cond.) give 3 eqns. for the normal balance (involving $R^{1,0,-1}$ terms), and 2 eqns. for the tangential balance (involving $R^{0,-1}$ terms).

These 11 resulting equations together with the condition (\ref{Qj}) provide a global system of 12 equations. However, only 8 of them are independent compatible algebraic equations, sufficient to solve the 8 unknown coefficients $\left. G_{j,i}^{jet}\right|_{j=2,3, i=1,2,3}$ and $\left. A_{3,i}^{jet}\right|_{i=1,2}$. However, the choice of the 8 equations has to be made carefully among all possible 8-element subsets of 12 elements. To do this, we follow this procedure:

1- First, solve the coefficients $\{G_{2,2},G_{2,3},G_{3,1},G_{3,2},G_{3,3}\}$ as functions of the remaining variables and unknowns by setting to zero on the jet surface the orders $R^1$ and $R^0$ of the normal velocity, the difference between the tangential velocities and the normal and tangential stresses. This yields just four independent homogeneous algabraic equations. The inhomogeneous equation remaining to complete the linear algebraic system of five equations is given by the order $R^0$ of the flow rate (\ref{Qj}).

2- Setting $A_{2,1}=A_{1,1}$ and $A_{2,2}=A_{1,2}$, solve $G_{2,1}$ by setting the order $R^{-1}$ of the tangential velocities on the jet surface to zero.

The orders of $R$ higher than $R^{-2}$ are null at this level after solving steps 1 and 2. Finally, three more equations come from setting the left order $R^{-3}$ of the tangential stress on the jet surface to zero. However, two of them are incompatible. Thus, leaving aside $R_J$, we solve the three possible compatible combinations of two equations among the three equations for the two unknowns $\{A_{3,1},A_{3,2}\}$. One of the choices minimizes the inhomogeneous part of the excluded incompatible equation. Using Mathematica$\circledR$, the procedure described finally produces the following solution:

\begin{eqnarray}
    G_{3,1} & = & -\frac{q}{\pi R_J^2}
    \left(\frac{3 \lambda \csc ^2\left(\frac{\alpha}{2}\right)}{\left(2 \lambda-1\right) \left(2 \cos \left(\alpha\right)+1\right)}+\frac{1}{2}\right) \label{G1}\\
    G_{3,2} & = &
    -\frac{q}{\pi R_J^2} \frac{6 \lambda \cot ^2\left(\frac{\alpha}{2}\right)}{\left(2 \lambda-1\right) \left(2 \cos \left(\alpha\right)+1\right)}\\
   G_{3,3} & = & \frac{q}{\pi R_J^2} \left(\frac{1}{2}-\frac{3 \lambda \cot ^2\left(\frac{\alpha}{2}\right)}{\left(2 \lambda-1\right) \left(2 \cos \left(\alpha\right)+1\right)}\right) \\
    G_{2,1} & = & -\frac{q}{\pi R_J^2} \left(\frac{3 \lambda^2 \cot ^2\left(\frac{\alpha}{2}\right)}{\left(2 \lambda-1\right) \left(2 \cos \left(\alpha\right)+1\right)}+\frac{1}{2}\right)\\
    G_{2,2} & = & -\frac{q}{\pi R_J^2} \frac{6 \lambda^2 \cot ^2\left(\frac{\alpha}{2}\right)}{\left(2 \lambda-1\right) \left(2 \cos \left(\alpha\right)+1\right)}\\
    G_{2,3} & = & \frac{q}{\pi R_J^2} \left(\frac{1}{2}-\frac{3 \lambda^2 \cot ^2\left(\frac{\alpha}{2}\right)}{\left(2 \lambda-1\right) \left(2 \cos \left(\alpha\right)+1\right)}\right) \label{G2}
\end{eqnarray}
\begin{eqnarray}
    A_{3,1} & = & q \frac{3 \csc ^4\left(\frac{\alpha}{2}\right) \left(\left(3-5 \lambda\right) \cos \left(2 \alpha\right)-3 \lambda+1\right)}{32 \pi  \left(2 \lambda-1\right) \left(2 \cos \left(\alpha\right)+1\right)}\\
    A_{3,2} & = & q \frac{\csc ^4\left(\frac{\alpha}{2}\right) \left(3 \left(\lambda-1\right) \cos \left(2 \alpha\right)+5 \lambda-1\right)}{32 \pi  \left(2 \lambda-1\right) \left(2 \cos \left(\alpha\right)+1\right)}.
\end{eqnarray}

So far, we have already obtained the best possible solutions to the surface problems of the cone (exact) and of the jet (exact up to the order $R^{-2})$). The next step is to finally solve the matching problem of solutions in regions (1) and (2) at an intermediate angle $\chi$ (see Figure \ref{regions}).

One might be tempted to consider solving the problem by linearizing the equations for small angles $\epsilon = R_J/R\ll 1$. However, this not only does not save any effort compared to using the full exact expressions, but also results in an unproductive mathematical burden compared to the optimized procedure described. Instead, an alternative complete formulation using a consistent slender approximation from the beginning is given in a separate section (see Section \ref{Jens}).

As already said, the outer stream functions $\psi_1^{(1)}$ and $\psi_1^{(2)}$ share the same part $A_{j,2}\cos ^3(\theta )+ A_{j,1}\cos (\theta )$ ($j=1,2$). However, the part multiplied by $R^2$, namely
\begin{equation}
\phi^{(j)}=G_{j,1}+G_{j,2} \cos (\theta ) + G_{j,3}\cos ^2(\theta ) + G_{j,4}\sin ^2(\theta ) \tanh ^{-1}(\cos (\theta ))
\end{equation}
is different at outside regions (1) and (2) due to the logarithmic part of the solution at region (1), which should be zero at region (2) to satisfy regularity at the axis when $R\rightarrow \infty$. A perfect match between both solutions is impossible for simple uniqueness reasons, but one can hypothesize the existence of a certain intermediate angle $\chi$ where the velocities and stresses can be matched. This would happen if the values of $\left.\phi^{(j)}(\theta)\right|_{j=1,2}$ and its first and second derivatives match at $\theta=\chi$.

Thus, the matching involves only the coefficients $G_{i,j}$, given by (\ref{G1})-(\ref{G2}) since the flow rate solution (coefficients $A_{i,j}$) is matched. This means that the quantity $U_J=q/(\pi R_J^2)$, that is, the average velocity of the jet, is the only relevant variable in the match, independently of the length scale or flow rate used. Without loss of generality, we set $q=1$. Solving the matching problem entails to find an angle $\theta=\chi$ mediating domains (1) and (2) where, defining
\begin{equation}
    \phi^{(1)}-\phi^{(2)}=\epsilon_0,\quad
    \partial_{\theta} \phi^{(1)}-\partial_{\theta} \phi^{(2)}=\epsilon_1,\quad
    \partial_{\theta}^2 \phi^{(1)}-\partial_{\theta}^2 \phi^{(2)}=\epsilon_2,\label{ec3}
\end{equation}
the solution to $\epsilon_i=0$ for $i=0,1,2$ would yield the exact matching of velocities and stresses at the intermediate angle $\chi$. Theoretically, once all coefficients of $\phi^{(j)}$ have been solved, this matching (three equations) would solve the three unknowns $R_J$ (which is the value of $R_J$ in units of the local scale $l_0$), the intermediate angle $\chi$ and the cone angle $\alpha$ for a given value of $\mu$. However, this theoretical {\sl exact} solution does {\sl not} exist because the three sheets defined by the equations (\ref{ec3}) in the space $ \{\alpha,\chi,R_J\}$ with $\epsilon_i=0$ do not meet at any non-trivial point (except at the trivial cylindrical solution $\alpha=0$, $\chi=\pi$ for any arbitrary large value of $R_J$). This is due to the failure of system (\ref{ec3}), defined by successive derivatives of the first equation, to satisfy the transversality conditions of Morse-Sard's theorem \cite{M39,S42} (i.e strict independency, or topological absence of parallelism or tangency of the manifolds defined by the equations) for $\epsilon_i=0$: in fact, derivation defines a linear relationship between equations (\ref{ec3}) that violates those independency conditions.

Nevertheless, the particular nature of system (\ref{ec3}), which is linear in $R_J$ and quadratic in $\lambda$, allows one to obtain an {\sl optimal} approximate solution of the matching problem. This optimal solutions yields a relationship for the matching angle of the form $\chi=\chi(\alpha;\lambda)$, given in the Appendix, such that the matching errors $\epsilon_i$ are strictly set to zero: although this does not result in unique $R_{J}$ values (by the failure to satisfy the conditions of Morse-Sard's theorem), one can find a relationship $\chi=\chi(\alpha;\lambda)$ that makes their differences strictly minimal. In fact, defining an error norm for their differences, the location $\xi=\chi(\alpha;\lambda)$ is graphically visualized as a narrow ``creek'' of that error (see Appendix A).

\begin{figure}
\centerline{\includegraphics[width=0.65\textwidth]{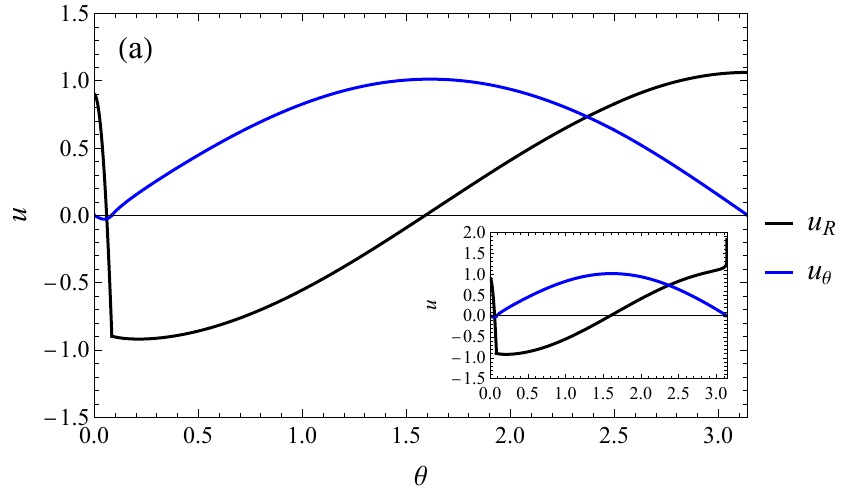}\includegraphics[width=0.50\textwidth]{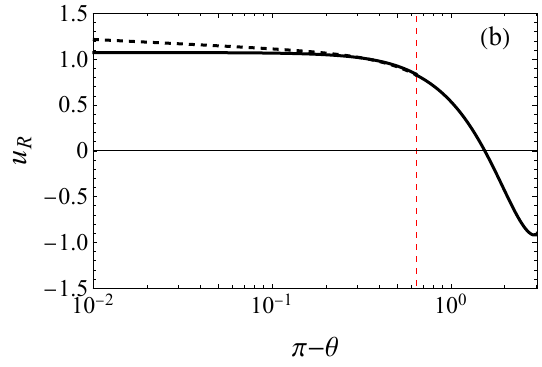}}
\centerline{\includegraphics[width=0.65\textwidth]{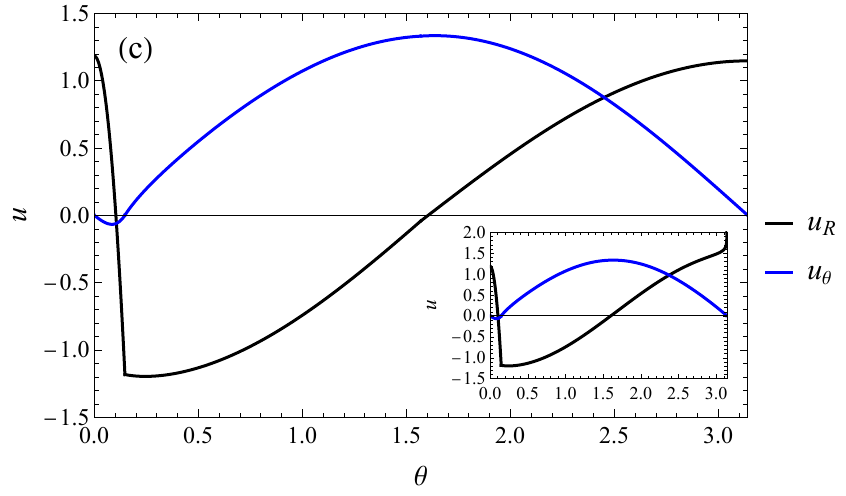}\includegraphics[width=0.50\textwidth]{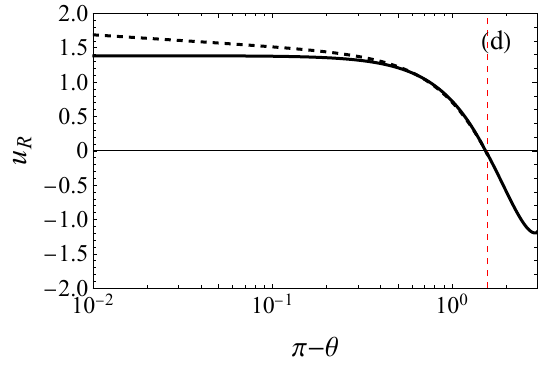}}
  \caption{Radial and angular velocities $u_R$ and $u_\theta$ for the analytical solution. The main plots show the solution regularized at the axis, while the insets show the exact solution (\ref{psol})-(\ref{As}) with a noticeable singularity at the axis ($\theta=\pi$). (a)-(b) $\chi = 2.5 \rightarrow \alpha = 0.0632$; (c)-(d) $\chi = \pi/2 \rightarrow \alpha = 0.1481$. $R=2000,\, \lambda=0.01$ in both cases. The matching angle $\chi$ values for given the $\alpha$ are calculated according to the functional relationship (\ref{fit}) in Appendix A. Here, the matching errors of $u_R$ ($\epsilon_2=0$) and $u_\theta$ ($\epsilon_1=0$) at $\theta=\chi$ are zero using the expression (\ref{jetradius}) for $R_{J,2}$ and $R_{J,1}$ respectively. In (b) and (d), a detail of the functional forms of the regularized (continuous black lines) and non-regularized (black dashed lines) is presented. The red vertical thin dashed lines represent the angular position where the approximate matching is performed.}
\label{f8}
\end{figure}

To illustrate how the cone jet solution comes from a regularization of the exact solution (\ref{gsol})-(\ref{As}), Figure \ref{f8} shows the radial and angular velocities $u_R$ and $u_\theta$ for large values $R$. In Figures \ref{f8}(a),(c), for $\lambda=0.01$, both the regularized (main plots) and the exact analytical solution with a logarithmic singularity at the axis $\theta=\pi$ (insets) are shown. In Figures \ref{f8}(b),(d) the functional form of the non regularized (denoted by dashed lines) and the regularized (continuous lines) solutions are plotted, showing the decreasing singular behaviour of $u_R$ for $\theta\rightarrow \pi$ for decreasing $\alpha$ values, that correspond to decreasing $\pi-\chi$ values too (see Appendix A).
To obtain these solutions, we have used the relationship between the cone angle $\alpha$ and the matching angle $\theta=\chi$ that minimizes the differences between the values of $R_J$ obtained when the matching errors for $u_R$ and $u_\theta$ are set to zero (see the Appendix A).

We emphasize that $\alpha$ {\sl is here a free parameter} under the assumption of a conical flow at infinity, with $\chi$ a function of $\alpha$ such that $\chi_{min} \lesssim  \chi <\pi$, such that $\chi_{min}\lesssim 1.22$ as shown in the  Appendix. However, the slender body theory in Section \ref{Jens} based on $\alpha\ll 1$, which admits other possibilities at infinity, shows that instead of $\alpha$ as a free parameter, a local {\sl physical} capillary number related to the strength of the external flow (not necessarily a conical flow) fixes $\alpha$.

%This result summarizes the ultimate reason for explaining the existence of almost conical tips with invisible ejections in the early physical observations of \cite{T34}, \cite{RM61}, the more recent experiments of \cite{GGRHF07}, and the numerical experiments of \cite{RMEH24} among several other works.

Nevertheless, for completeness and consistency, our proposed local analytic cone-jet solution is completed by the numerical solution at the cone-jet transition region, which is tackled next.

% Numerical
\subsection{The cone-jet intermediate region: numerical implementation}

\begin{figure}
\centering
\includegraphics[width=0.60\textwidth]{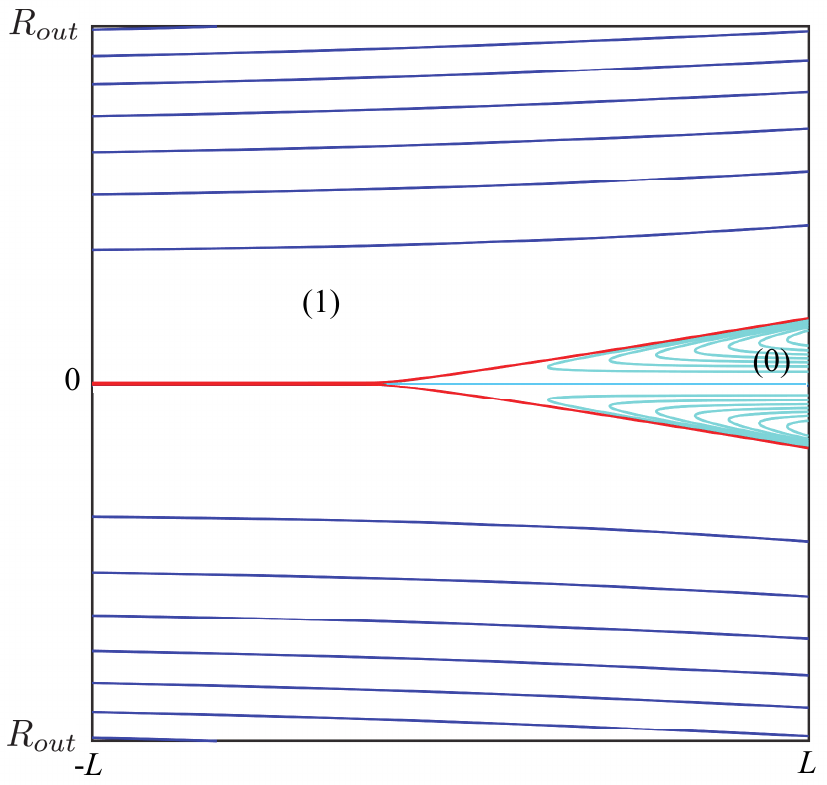}
\caption{Numerical solution of the intermediate cone-jet region for $\lambda = 0.0125$ and $\alpha=0.173$, using the asymptotic analytic cone-jet solution as the far boundary conditions of domains (0) and (1), except the outlet conditions of the jet specified in the text. The flow pattern is virtually identical to that of the analytical solution (\ref{Gs})-(\ref{A3}) with $q=0$ used in Figure 1.}
\label{f5}
\end{figure}

For a given set of $\lambda$ and $\alpha$ values, the intermediate region between the cone, a perfectly conical recirculating flow with a nonzero net flow rate (i.e. $q=1$), and an asymptotically cylindrical jet with a radius at infinity proportional to the local scale will be solved numerically using a cylindrical coordinate system ($z$, $r$), with $z=R \cos \theta$ and $r=R \sin \theta$, in a rectangular domain $[-L,L]\times[0,R_{out}]$ sketched in figure \ref{f5}.

The numerical code solves the conservation of mass and a balance of linear momentum in the outer ($j=1$) and inner ($j=0$) subdomains, given by
\label{eq:gov}
\begin{align}
\label{equ1}
\grad\cdot{\bf v}_j &= 0, & &(j=0,1),\\
\label{equ2}
\grad p_{j}&=\lambda_{j}\grad^2 \bf v, & &(j=0,1),
\end{align}
where  ${\bf v}_j=w_j \mathbf{e}_z+ u_j\mathbf{e}_r$  is the velocity field, $p_j$ is the pressure, and $\lambda_1=1$ as previously stated. We use the analytical solution as the far-field boundary conditions at the computational box boundaries:

\begin{enumerate}
\item At right boundary, $z= L_{box}$, we assume  that the numerical solution matches the analytical solution for the cone,
\beq
w_j=\frac{1}{r}\frac{\partial \psi_i^{cone}}{\partial r} \quad u_{j}=-\frac{1}{r}\frac{\partial \psi_j^{cone}}{\partial r}, \quad (i=0,1),\label{entranceflow}
\eeq
and $F=L_{box}\tan(\alpha)$, where $r=F(z,t)$ is the  parametric representation of the interface in terms of $z$ (see margenta line in figure \ref{f5}).

\item At the  left boundary, $z=-L_{box}$,  we impose the following Neumann condition for the outer flow
\beq
\frac{\partial w_1}{\partial z}=\frac{\partial w_1^{jet}}{\partial z},\quad\frac{\partial u_1}{\partial z}=\frac{\partial u_1^{jet}}{\partial z},
\label{outflow1}
\eeq
where $w_1^{jet}$ and $u_1^{jet}$ are the velocities obtained with the analytical stream function of the jet solution ($\psi_1^{jet}$). On the other hand,  for inner flow, outflow conditions are considered
\beq
\frac{\partial w_0}{\partial z}=0,\quad u_0=0, \quad \frac{\partial F}{\partial z}=0.
\label{outflow2}
\eeq

\item At the  top boundary, $r=R_{out}$, The velocity varies continuously from the analytical solution associated with the cone to the solution associated with the jet.
\beq
z<-\sin(\chi):\quad w_1=w_1^{jet} , \quad z>=-\sin(\chi): w_1=w_1^{cone}\quad  u_1=u_1^{cone}.
\label{top}
\eeq

\end{enumerate}

Across the gas-liquid interface, we use the same conditions (\ref{normal})-(\ref{un}) already expressed for the analytical solution.

% Numerical method.
The governing equations are integrated with a variant of the numerical method proposed by \cite{HM16a,JAM}. The  inner (0) and outer (1) domains are mapped onto rectangular domains by means of the analytical mappings:
\beq
r=F(\xi,t)\zeta_0,\quad z=\xi,\quad [0\leq \zeta_0\leq 1]\times [-L\leq \xi\leq L]%[-L_1{/h}\leq \xi_1\leq {(L+L_2)/h}]\times[0\leq \zeta_g\leq 1],
\eeq
for the inner domain, and
\beq
r=F(\xi,t)+\zeta_1[R_{out}-F(\xi,t)],\quad z=\xi,\quad [0\leq \zeta_0\leq 1]\times [-L\leq \xi\leq L].%[-L_1{/h}\leq \xi_1\leq {(L+L_2)/h}]\times[0\leq \zeta_g\leq 1],
\eeq
for the outer domain. These mappings are applied to the governing equations, and the resulting equations are discretised in the $\zeta$-direction with $n_{\zeta_0}$ and $n_{\zeta_1}$ Chebyshev spectral collocation points in the inner and outer domains, respectively.   Conversely, in the $\xi$-direction we use second-order finite differences with $n_{\xi}$ points.

%Basic flow
Steady-state solutions of the nonlinear discretized equations  with all variables independent of time  are obtained by solving all equations simultaneously (a so-called monolithic scheme) using a Newton-Raphson procedure.

The analytical numerical cone-jet solution completed so far will be simply called the ``cone-jet solution''. To summarize the physical framework of this solution, we have:

1- It implies a self-similar conical flow at infinity.

2- The intermediate cone-jet solution is resolved assuming $q=1$ This means that, for numerical simplicity purposes, the intermediate scale $l$ is assumed as $l=l_o$.

3- Therefore, from the previous points, the intermediate length $l$ should be very small compared to any other macroscopic length, including that for which a nearly conical flow is observed.

4- Thus, using $l_o$, $m_o$ and $t_o$ as the units of length, mass and time, we also have the dimensional value of the flow rate $Q=1$.

5- This solution has two independent variables: the viscosity ratio $\lambda$ and the cone angle $\alpha$. The latter is not restricted except for that there is a maximum velocity that the cone surface can attain, and this is obtained for a specific value of the angle: $\alpha_m = 2 \lambda^{1/2}$.

Once both the cone-jet solution and the slender body solution proposed in the following are obtained, their comparison will shed light on the long-standing nontrivial problem they resolve and will explain the existence of almost conical tips with invisible ejections in the early physical observations of \cite{T34}, \cite{RM61}, the more recent experiments of \cite{GGRHF07}, and the numerical experiments of \cite{RMEH24} among several other works.

\section{Slender body description of the transition region}
\label{Jens}

\begin{figure}
\centerline{\includegraphics[width=0.99\textwidth]{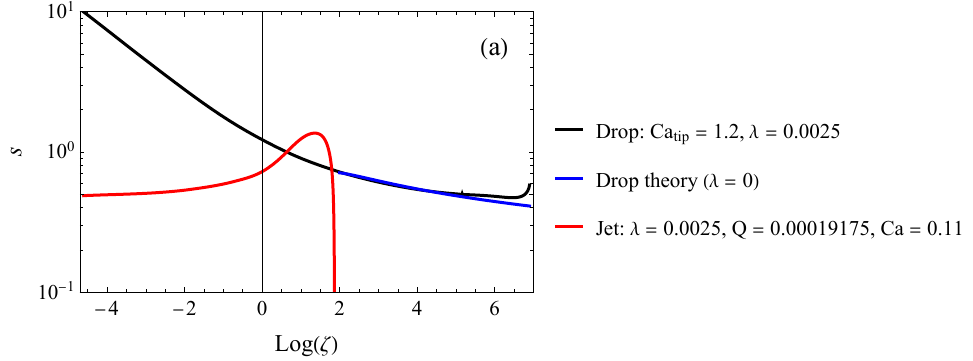}}
\centerline{\includegraphics[width=0.99\textwidth]{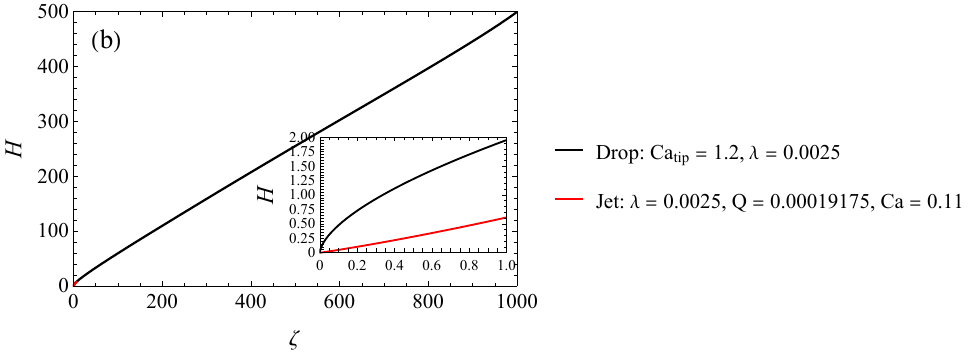}}
  \caption{(a) Slopes $s$ of strained menisci by an extensional outer flow. "Drop": non-emitting tip-rounded menisci, \cite{E21}; "Jet": meniscus with emission, with $Q \rightarrow 0$. The black and red lines have been obtained numerically, while the blue line is the theory of \cite{E21}. The parameter $Ca_{tip}$ is the capillary number defined with the scaling of the maximum curvature $\kappa_{m}$. (b) Scaled profiles $H(\zeta)$ of the "Drop" and "Jet" menisci obtained numerically. Here, $\zeta$ is either the axial coordinate $z$ scaled with $\kappa_m$ ("Drop") or with the outer scale, in this case the tube radius $R_0$ of Figure \ref{f1} ("Jet").}
\label{f6}
\end{figure}

The scaling \eqref{lambda_scale} of the cone solution \eqref{psol} suggests that in the limit of small $\lambda$, slopes are small, and an approximation based on the slenderness of the jet should be applicable \cite{B72,T66b,Z04,CCG12}. We follow \cite{Z04} in applying Taylor's slender body theory, developed originally to study drops and bubbles of small or vanishing viscosity in an extensional flow, to jetting solutions. While slender body theory is known to fail near conical drop tips \cite{T66b,B72,E21}, our comparison of Figure \ref{f6} indicates that this does not apply in the jetting case, even as the jet diameter goes to zero.
In fact, the comparison between the local profiles of non-emitting conical drop tips and jetting conical tips (Figure \ref{f6}(b), inset) reveals the totally different nature of both solutions.

\subsection{Slender body theory}

We now show that Taylor's theory, developed for driving by an extensional flow, can be applied to an arbitrary external flow ${\bf u}^{(ext)}(z,r),p^{(ext)}(z,r)$, which we assume satisfies the Stokes equation in cylindrical coordinates. In the slender limit, the jet is expected to make a small correction, so we write the flow in the outer phase as
\[
{\bf u}^{(out)} = {\bf u}^{(ext)} + {\bf u}',
\]
where ${\bf u}'$ is a distribution of two-dimensional point sources along the axis, with strength $A(z)$: ${\bf u}' = (0,A(z)/r)$. This approximation can be verified by a systematic calculation based on an expansion in the slenderness \cite{B72}, but physically it means that the jet's only effect on the outer flow is to occupy extra volume. Now $A(z)$ follows from the kinematic boundary condition ${\bf u}\cdot{\bf n} = 0$:
\beq
A(z) = h\left[u_z^{(ext)}(z,h)h' - u_r^{(ext)}(z,h)\right].
\label{A}
\eeq

In the interior, using slenderness, we can assume a parabolic lubrication-type flow, while in the limit of small $\lambda$ the external flow is unchanged; hence we obtain
\beq
u^{(in)}_z = -\frac{1}{4\lambda\mu_0}\frac{dp}{dz}(h^2-r^2) + u_z^{(ext)}(z,h),
\label{u_int}
\eeq
using continuity of the velocity. The shear balance does not need to be taken into account to leading order \cite{B72}, and the normal force balance reads for a slender thread:
\beq
\sigma_{rr}^{(out)} + p^{(in)} = \frac{\gamma}{h},
\label{kappa}
\eeq
using the leading order expression for the mean curvature. Calculating the stress, we obtain
\begin{eqnarray}
\sigma_{rr}^{(out)} = -p^{(ext)} + 2\mu_0\frac{\partial u_r^{(ext)}}{\partial r}
- \frac{2\mu_0 A(z)}{h^2} =\nonumber \\
-p^{(ext)} +
2\mu_0\frac{\partial u_r^{(ext)}}{\partial r}
- \frac{2\mu_0}{h}\left[u_z^{(ext)}(z,h)h' - u_r^{(ext)}(z,h)\right].
\end{eqnarray}

Expanding in the slenderness, we have $u_z^{(ext)}(z,h) = u_0(z) + O(h^2)$
and $u_r^{(ext)}(z,h) = -u_0'(z)h/2 + O(h^3)$, using incompressibility.
Thus to leading order we finally obtain for the inner pressure $p^{(in)}\equiv p$:
\beq
p = \gamma\kappa + p_0(z) + \frac{2\mu_0}{h}(h(z)u_0(z))',
\label{p_lub}
\eeq
where $u_0(z)$  and $p_0(z)$ are the values of $u_z^{(ext)}$ and $p^{(ext)}$ on the axis, respectively. In the interior, in a steady state, the flux is constant:
\beq
\frac{Q}{\pi} = -\frac{h^4}{8\lambda\mu_0}\frac{dp}{dz} +
h^2 u_0(z).
\label{Q_lub}
\eeq
Thus \eqref{p_lub} and \eqref{Q_lub} are a closed set of equations for $h(z)$, provided that $u_0(z),p_0(z)$, which characterize the outer flow, are provided.

\subsection{Similarity solutions of the transition region}
\label{Similarity}

We are interested in a local description of the entry into a thread, taken to be at $z_0$, which is unknown, and around which we expand. As in section~\ref{sub:local}, $\sqrt{\mu_0 Q/\gamma}$ is a length scale. This suggests
the similarity solution
\beq
h = \sqrt{\frac{Q\mu_0}{\pi \gamma}} \lambda^{1/4}H(\xi), \quad
\xi = \sqrt{\frac{\pi \gamma}{Q\mu_0}}\lambda^{1/4} (z-z_0), \quad
\overline{\rm Ca} = \frac{u_0(z_0)\mu_0\lambda^{1/2}}{\gamma},
\label{sim}
\eeq
where $\overline{\rm Ca}$ is a local capillary number scaled as suggested by \eqref{US}, \eqref{lambda_scale}.
Since the local length scale vanishes in the limit of $Q\rightarrow 0$, $u_0(z)$ and $p_0(z)$ can be expanded about $z_0$, and only $u_0(z_0)$ matters at leading order. Moreover, for small $\lambda$, the slope scales like $-h' =\alpha \propto \lambda^{1/2} \ll 1$, so the profile is flat in this limit, and the slender jet approximation can be applied.

Inserting \eqref{sim} into \eqref{p_lub} and \eqref{Q_lub}, and retaining only the leading terms of order $Q$, in the limit of small $\lambda$ one finds
\beq
1 =  \overline{\rm Ca} H^2 + \frac{H^2H'}{8} +
\frac{\overline{\rm Ca}}{4}\left(H^2H'^2-H^3H''\right),
\label{ss_H}
\eeq
which is the similarity description valid for $\lambda \ll 1$. For $\xi\rightarrow\infty$ (the thread), \eqref{ss_H} has a constant solution of the form
\beq
H = \frac{1}{\sqrt{\overline{\rm Ca}}}, \quad
\xi\rightarrow\infty,
\label{ss_thread}
\eeq

On the other hand for $\xi\rightarrow-\infty$, \eqref{ss_H} allows for two linear solutions of the form $H = -s\xi$, where
\beq
s_{\pm} = \frac{1}{4\overline{\rm Ca}}
\left(1\pm\sqrt{1-(8 \overline{\rm Ca})^2}\right),
\label{ss_slope}
\eeq
as long as $\overline{\rm Ca}\le 1/8$. Taking $\overline{\rm Ca}$ as given, there is no free parameter; thus we aim to solve \eqref{ss_H}, with $\overline{\rm Ca}$ as a parameter. This situation is shown on Fig.~\ref{f8}(a), for which
a slope of $s_- = 1$ is approached.
\begin{figure}
\centering
\includegraphics[width=0.95\textwidth]{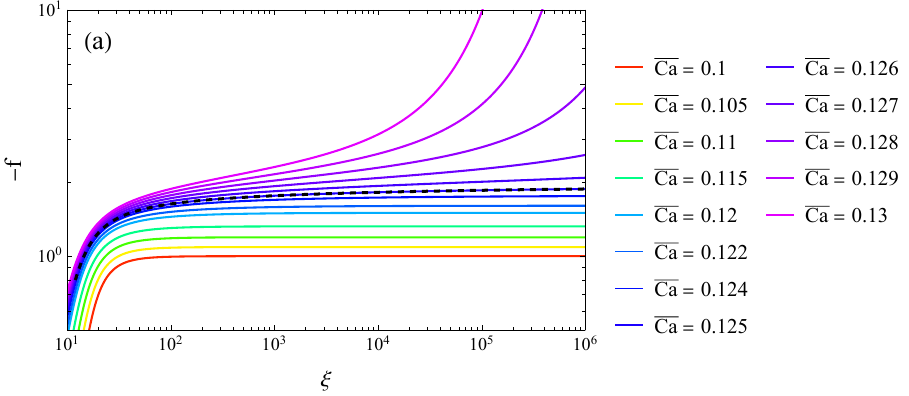}
\caption{
Non-dimensional scaled slope $-f$ of the cone-jet profile, for different $\overline{\rm Ca}$ values and $s_-$. The black dashed line underlines the $\overline{\rm Ca}=1/8$ case.}
\label{f7}
\end{figure}
To do this, we reduce the order by putting $f(H) = H'$, giving
\beq
f' = \frac{1}{fH}\left(\frac{f}{2\overline{\rm Ca}}+4+f^2\right)
- \frac{4}{\overline{\rm Ca}f H^3} =
\frac{1}{fH}\left(f+s_-\right)\left(f+s_+\right)
- \frac{4}{\overline{\rm Ca}f H^3}.
\label{ss_f}
\eeq
To find the profile $H(\xi)$, we solve simultaneously $\xi'(H) = 1/f$; the constant of integration is an arbitrary shift of the origin, which we chose as the point of maximum curvature. We solve \eqref{ss_f} starting from the
neighborhood of $H = H_m = 1/\sqrt{\overline{\rm Ca}}$:
\[
H = H_m + \delta, \quad
f = -\frac{-\sqrt{\overline{\rm Ca}} +
\sqrt{128 \overline{\rm Ca}^3} + \overline{\rm Ca}}{4\overline{\rm Ca}}\delta
\]
for small $\delta$. As a result, the solution is determined uniquely.

In Fig.~\ref{f7} we illustrate how only the smaller root $s_-<s_+$ is selected, it being approached by $-f$ from below. Namely, the second contribution on the right of \eqref{ss_f} is always positive, while the first vanishes as $s_-$ is crossed. Thus $f'$ is positive, so $-f$ can only decrease, making it impossible to cross. On the other hand, for
 $\overline{\rm Ca} > 1/8$ there is no more linear solution. Instead, the leading balance in \eqref{ss_f} is now $f' \approx f/H$, because the term $f^2$ in brackets on the left dominates. Thus $f = f_0 H$, and integrating we have $H = H_0 e^{f_0\xi}$, the profile is increasing exponentially (see Figure \ref{f7}). Consequently, despite the fact that the approach to this behavior is extremely slow, so that it looks like a linear profile with a slowly varying slope, the profile would never reach the $Q\rightarrow 0$ limit with arbitrary but finite boundary conditions, a limit that is conceptually achievable by the conical profile.

So far, $\overline{\rm Ca}$ is a parameter which can take any value, depending on the position of $z_0$. Indeed, since $u(z)$ is expected to vary along the axis, $\overline{\rm Ca}$ can in principle take any value. It must adjust itself as part of matching of the local profile to the global solution. One possibility is that the solution adjusts itself such that $\overline{\rm Ca} \approx 1/8$. Intuitively, in a regime where the flow strength is increasing with $z$, this might be understood by the following argument: if $z_0$ moves forward, $\overline{\rm Ca}$ increases, pushing $z_0$ further forward. This will continue up to reaching a point where the similarity (conical) solution can no longer be matched, which is only possible as long as the profile is linear. This suggests that the similarity solution is tuned to its critical value. This suggests that $z_0$ is selected such that
\beq
\frac{u_0(z_0)}{v_c} \approx \frac{1}{8\sqrt{\lambda}},
\label{slope_cr}
\eeq
in the limit of small $Q$. This would mean that the local capillary number $\overline{\rm Ca}$
%becomes large
is maximized for small $\lambda$, and in turn the local slope of the conical region is $h' =-2\sqrt{\lambda}$, in remarkable coincidence with (\ref{lambda_scale}): it is equal to the cone angle $\alpha_m$ that maximizes the local speed on the cone surface for the initial exact analytical solution (\ref{psol})-(\ref{As}). Since this slope is small in the limit $\lambda\rightarrow 0$, this means that the slender body approximation is justified in this limit.

In summary, compared to the cone-jet solution, the physical framework of this slender body solution implies the following:

1- This solution has two independent variables: the viscosity ratio $\lambda$ and the local capillary number $\overline{\rm Ca}=\frac{u_0(z_0) \mu_0\lambda^{1/2}}{\gamma}$, where $u_0(z_0)$ is the velocity of the external flow on the axis at the location $z=z_0$, in the absence of a jet as a first approximation: It measures the strength of the external flow.

2- The cone angle $\alpha$ of the cone-jet solution is related to $\overline{\rm Ca}$ of the slender body solution by the expression
\beq
\alpha=\frac{1-\sqrt{1-(8\overline{\rm Ca})^2}}{4\overline{\rm Ca}}\lambda^{1/2}.
\label{alphaCa}
\eeq
%However, while the analytical solution does not see other conditions at infinity than a perfect conical flow with the cone angle $\alpha$ as a free parameter, the slender body theory can physically relate its parameter, the local capillary $\overline{\rm Ca}$, with the strength of the macroscopic driving flow.

3- The latter is not restricted except for that there is a maximum value $\overline{\rm Ca}_{cr}=1/8$ to have a conical flow at infinity. This $\overline{\rm Ca}_{cr}$ leads to a slope that coincides with the angle of the cone jet solution for which the velocity on the cone surface is maximized: $\alpha_m = 2 \lambda^{1/2}$.

4- It resolves the complete cone-jet intermediate region. It also provides a {\sl universal} solution in terms of rescaled variables that involve the properties of the liquid, the liquid flow rate, and the local strength of the external flow. Naturally, this strength can be related to the surface velocity in the cone through a universal constant of the order unity that depends on the external velocity field used.

5- The flow rate $Q$ can be obtained in terms of the value of $\overline{\rm Ca}$, the liquid properties, and the velocity of the external flow. Like in the cone-jet solution, if one uses $l_o=\sqrt{\mu_0 Q/\gamma}$ as the length scale, $Q=1$.

\section{Results}

In this section we present examples using the cone-jet and the slender body solutions. First, we compare both solutions at the local scale $l_o$. Afterwards, for consistency, we give examples of the comparison of the slender-body solution with complete simulations with near-real boundary conditions, which underscores critical aspects revealed by the slender-body solution.

\subsection{A local comparison of cone-jet and slender-body solutions}

\begin{figure}
\centerline{\includegraphics[width=0.95\textwidth]{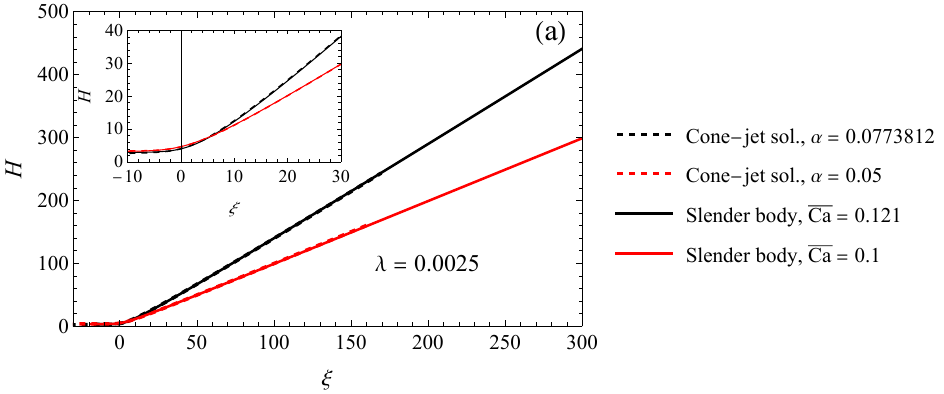}}
\centerline{\includegraphics[width=0.95\textwidth]{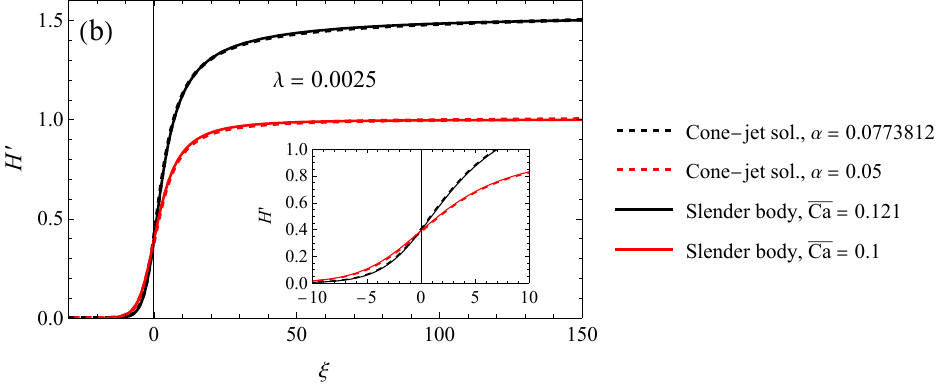}}
  \caption{Comparison of the cone-jet and slender body solutions for $\overline{\rm Ca}<1/8$, or alternatively $\alpha < \alpha_m$ for two $\overline{\rm Ca}$ values. Here, $\lambda=0.0025$; $\alpha$ and $\overline{\rm Ca}$ are related by (\ref{alphaCa}). (a) The cone-jet profiles. (b) The cone-jet slopes.}
\label{f9}
\end{figure}

Secondly, in Figure (\ref{f9}) we compare the cone-jet and the slender body solutions for $\overline{\rm Ca}<1/8$, which corresponds to $\alpha < \alpha_m$. The relationship between $\overline{\rm Ca}$ and $\alpha$ is given by (\ref{alphaCa}). After rescaling the cone-jet solution according to the definitions (\ref{sim}), the agreement is nearly perfect, as expected since $\lambda \ll 1$.

In general, while the slender body theory assumes that a consistent error of the order of $O(\alpha^n)$ with $\alpha \ll 1$ and $ n > 1 $ occurs and affects all variables everywhere, the cone-jet solution assumes that the errors are restricted to the region where two solutions of the external flow are matched, imposing exact conditions at the cone and the jet. In contrast, the slender body theory uses a physically meaningful parameter $\overline{\rm Ca}$, resolves the cone-jet region, reduces the problem to a universal scaling structure of the cone-jet geometry, and indicates the maximum value of the local flow strength for which the flow can be conical for a given viscosity ratio $\lambda$.

\subsection{Comparison of the slender body solution with a complete numerical solution.}

\begin{figure}
\centerline{\includegraphics[width=0.70\textwidth]{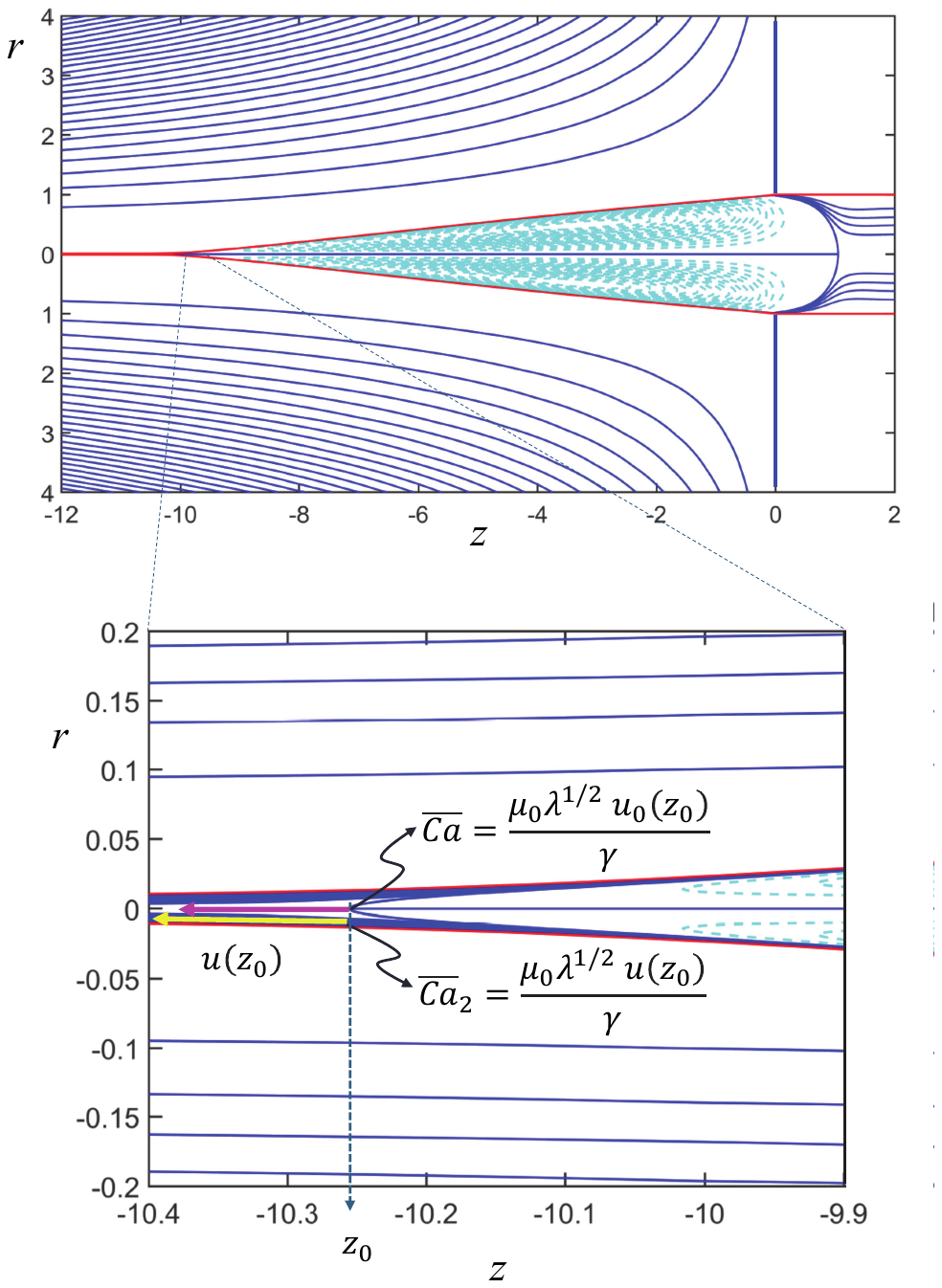}}
  \caption{Shape and streamlines of a cone-jet meniscus subject to a purely extensional flow \cite{RMEH24}. The definition of the local capillary number $\overline{\rm Ca}_2$ is indicated, where instead of the velocity of the external flow at the stagnation point $z_0$ on the axis, $u_0(z_0)$, we use the velocity of the interface of the cone-jet at the same $z_0$, i.e. $u(z_0)$. The magenta arrow indicates the value of the velocity of the external flow at the axial location of the stagnation point of the inner flow, in the absence of the latter, necessary to calculate $\overline{\rm Ca}$ as defined, while the yellow arrow indicates the value of the velocity on the interface at the axial location $z=z_0$ to calculate $\overline{\rm Ca}_2$.}
\label{f10}
\end{figure}

In this section, we provide an illustration of the hint anticipated by the end of section \ref{Similarity}, suggesting that the existence of a critical $\overline{\rm Ca}_{cr}$ implies that, for increasing strengths of the external flow over a certain value which depends on the nature of that flow, the local flow is tuned to its critical value. To this end, we use the extensional flow solution used in \cite{RMEH24}; see Figure \ref{f10} for details of the definitions used to perform the comparison.

\begin{figure}
\centerline{\includegraphics[width=0.95\textwidth]{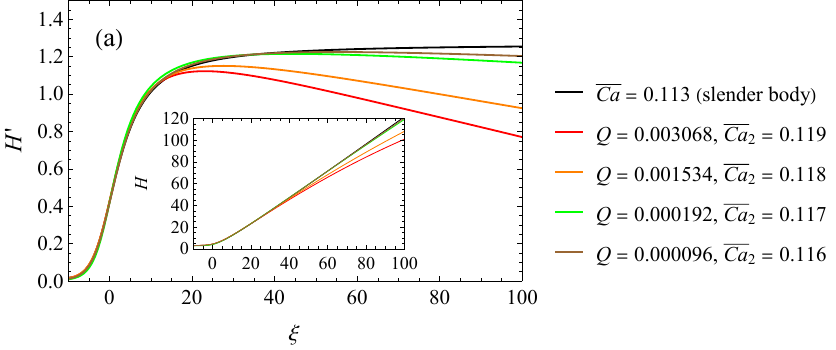}}
\centerline{\includegraphics[width=0.95\textwidth]{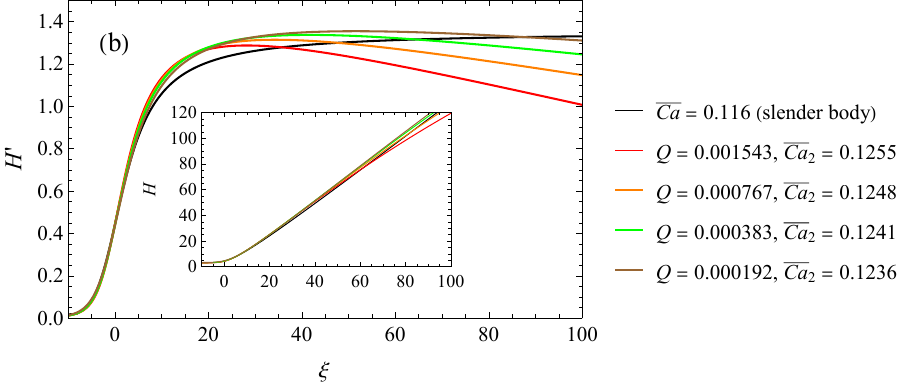}}
\centerline{\includegraphics[width=0.95\textwidth]{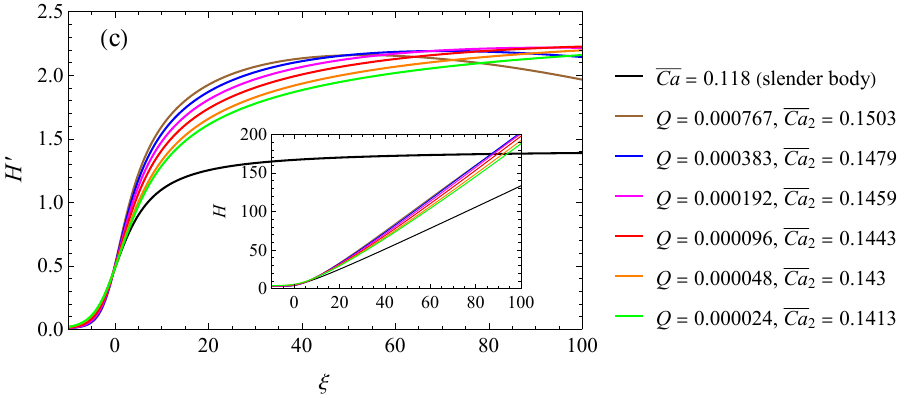}}
  \caption{Normalized cone-jet shapes $H(\xi)$ and their slopes $H'(\xi)$ for $\lambda = 0.025$ and three different $C$ values: (a) $0.09$, (b) $0.11$ and (c) $0.3$. Black lines correspond to the slender body theory, while color lines correspond to the numerical case with actual boundary conditions and different flow rates \cite{RMEH24}.}
\label{f11}
\end{figure}

\begin{figure}
\centerline{\includegraphics[width=0.8\textwidth]{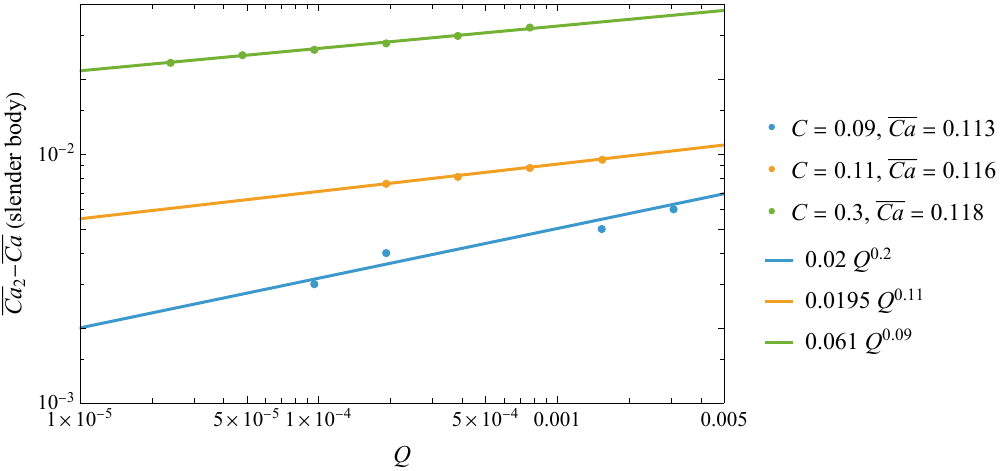}}
  \caption{Convergence of the ``external'' local capillary number $\overline{\rm Ca}_2$ of the numerical solution to the slender body capillary number $\overline{\rm Ca}$ as $Q\rightarrow 0$, for the three cases of Fig.~\ref{f11}}
\label{f12}
\end{figure}

Figure \ref{f11} represents the local cone-jet transition from the complete numerical solution of a flow according to Figure \ref{f10}, compared to the solution provided by the slender-body theory. To provide the most consistent comparison, we define a local capillary $\overline{\rm Ca}_2$ calculated with the velocity of the flow at the interface of the complete solution under the same extensional flow, at the same axial location $z=z_0$ as the definition of $\overline{\rm Ca}$. The value of $\overline{\rm Ca}$ of the slender body solution is obtained by matching the value of the rescaled profiles and their first and second derivatives at $z=z_0$, with the complete solution in the case of the lowest possible flow rate value $Q$ achievable by our numerical method. We have used three values of the external macroscopic capillary number $C=U \mu_0/\gamma$ of the extensional flow (where $U$ is the external characteristic velocity, \cite{RMEH24}), namely $C=0.09$, $0.11$ and $0.3$. To
find the values of $\overline{\rm Ca}$ used in the slender body description, we extrapolate
$\overline{\rm Ca}_2$ as found from full numerical solutions for successively smaller $Q$, as
shown in Fig.~\ref{f12}. In each case, $\overline{\rm Ca}_2$ converges like a power law toward  a limiting value of $\overline{\rm Ca}_2$. However, the power law exponent decreases with increasing $C$, indicating progressively slower convergence. Given the $\overline{\rm Ca}$-value thus found, the convergence of the numerical solutions to the slender body theory is excellent as $Q\rightarrow 0$ for the smaller $C$ value.
%in fact, given any small but finite flow rate $Q$, the slope can be matched at intermediate ranges of $\xi$ for specific values of $\overline{\rm Ca}>1/8$, while the slope at $\xi=0$ appears to correspond to slender body solutions with $\overline{\rm Ca}=1/8$.
Intriguingly, the local solution around $z=z_0$ appears to be below $\overline{\rm Ca}_{cr}$ for the three values of $C$ used in this study, although the convergence is slower as $C$ increases.
All this is illustrated in Figure \ref{f12}.

In conclusion, our results suggest that when the external capillary number is small, the rescaled macroscopic solution converges relatively rapidly to the slender body solution at large scales as $Q$ decreases. However, as the external flow strength increases, the local flow appears to approach $\overline{\rm Ca}_{cr}=1/8$ from below, and the corresponding local angle $\alpha$ of the cone tends to $\alpha_m=2\lambda^{1/2}$, but the approach of the numerical solution to the macroscopic cone angle $\alpha$ eventually seems to demand extremely small values of $Q$.

\subsection{Local universality adding far-field scales: numerical solutions with non-conical macroscopic boundary conditions}

Since the local solutions presented correspond to imposed macroscopic flow configurations, and these configurations are generally determined by a parameter that reflects the {\sl strength} of the external flow, this strength should be also reflected by an {\sl internal} parameter. The emergence of local universality is demonstrated using slender body theory. This local solution lacks any macroscopic far-field scale, except a local capillary number $\overline{\rm Ca}$ as proposed. The introduction of far-field scales and parameters can then be carried out numerically, as in a recently published configuration \cite{RMEH24}, where the authors did not attempt to find any local universality. However, when such results are examined for the smaller possible emitted fluxes compatible with the numerical code, the local universality here found emerges naturally.

From an analytical perspective, the geometry of the far-field external flow may involve velocity field expansion components in spherical coordinates such as $R^{n}$ with $n > 2$ to verify boundary conditions at far-field scales. Since we hypothesize that these components disappear at the intermediate length $l$, for scales larger than this intermediate scale where our solution is valid, an augmented stream function can be expressed in spherical local coordinates as
\begin{equation}
    \Psi^{(1)}=\frac{G_1}{2}R^3 \sin^2\theta \cos \theta.
\end{equation}
The $R^3$ behavior of this solution causes an unavoidable deviation in the far field from the cone. However, even in this case, the local universality shown by our local conical solutions still emerges when the emitted flow rate is numerically decreased to the method's working limits.

\section{Discussion and conclusions}

The three approximations presented in this work (i.e. analytical, slender body theory, and numerical) show excellent agreement when the local flow admits a strictly conical solution, which according to slender body theory occurs for local capillary numbers $\overline{\rm Ca}<1/8$. This imposes a strict maximum value of the axial velocity of the external flow at the tip of the cone in the absence of the cone, above which the local flow cannot be conical, but rather cusp-like.

The limit set by the local capillary number $\overline{\rm Ca}=1/8$, which corresponds to the value of the cone angle $\alpha_m$ for which the analytical solution has a maximum velocity at the interface, represents a strong conceptual separation. The conical geometry ($\overline{\rm Ca}<1/8$) allows a quick approach of the flow rate $Q$ to zero from any arbitrary macroscopic flow scale, as long as the strength of the macroscopic flow is weak (i.e., small macroscopic capillary values $C$). This is due to the self-similar conical geometry of the flow at any vanishing intermediate scale. However, any value $\overline{\rm Ca}$ above $1/8$, corresponding to a cusp geometry, while apparently allowing a local vanishing flow rate, cannot be assimilated to an arbitrary intermediate scale of vanishing size compared to the macroscopic scales, and therefore the flow rate cannot be made arbitrarily small. In fact,
the numerical solution suggests that for strong enough macroscopic flows, the local flow will eventually converge to the local conical solution with $\overline{\rm Ca}=1/8$, being it the {\it only one} possible local solution for arbitrarily strong macroscopic flow conditions. More detailed analysis is required to determine this.

In addition, we used an external extensional flow as the macroscopic flow configuration in this work as a bench test. Because the acceleration at the axis of this flow is very weak for axial coordinates beyond unity, the numerical solution shows very elongated menisci. Other configurations that create stronger local flow convergence (like flow focusing or selective withdrawal) will allow shorter menisci. Selective withdrawal may result in a unique solution for a given $\lambda$ ratio of viscosities, resulting in a conical tip with a specific value of the macroscopic $C$ matching the extreme local conical solution $\overline{\rm Ca} = 1/8$. This will be a topic of future analyses.

\section*{Declarations}

\begin{itemize}
\item Funding: Partial funding from the the Spanish Ministry of Science and Innovation, grant no. PID2022-14095OB-C21.
\item Conflict of interest/Competing interests (check journal-specific guidelines for which heading to use): The author declares no conflicts of interest.
\item Ethics approval: Not applicable
\item Consent to participate: Not applicable
\item Consent for publication: Not applicable
\item Availability of data and materials: Data and materials can be made available upon reasonable request.
\item Code availability: Not applicable
\item Authors' contributions: JE: Slender model: Conception, calculations and writing. MAH: Numerical model: Conception, calculations, writing, and funding. AMGC: paper conception; analytic model: conception, calculations, writing and funding.
\end{itemize}

\appendix

\section{Approximate analytical solution}

\begin{figure}
\centerline{\includegraphics[width=0.50\textwidth]{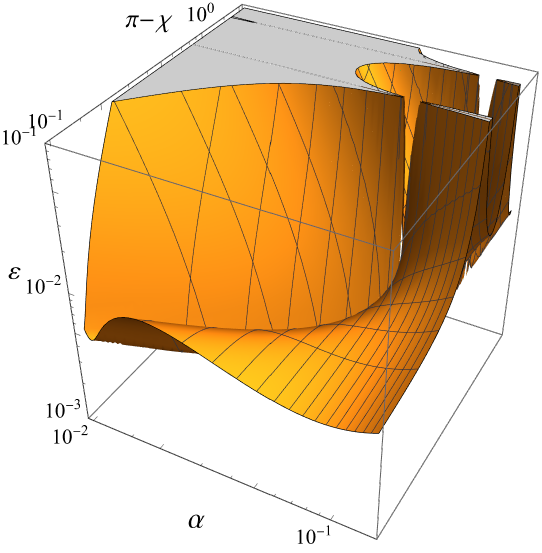}}
  \caption{The norm error between the values of $R_J$ for zero matching errors $\epsilon_i=0$. Here, $\lambda=0.005$.}
\label{f13}
\end{figure}

\begin{figure}
\centerline{\includegraphics[width=0.70\textwidth]{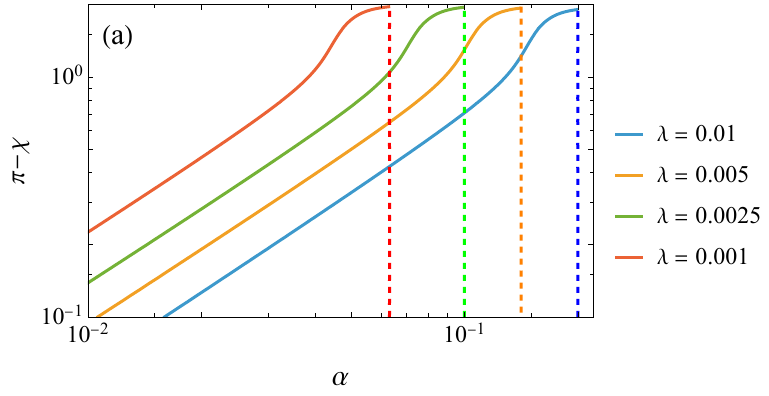}}
\centerline{\includegraphics[width=0.75\textwidth]{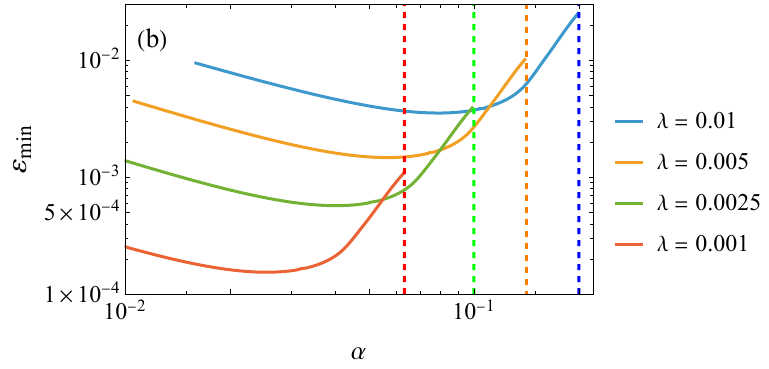}}
  \caption{(a) Locations of the local minima $\varepsilon_{min}$, in $\{\alpha,\pi-\chi\}$ and (b) values of the minima, for different $\lambda$ values. Dashed lines indicate the $\alpha_{max}$ value of the cone angle for each $\lambda$.}
\label{f14}
\end{figure}

In principle, the three equations (\ref{ec3}) define a system of three equations with three unknowns $\{R_J,\alpha_o,\chi\}$. Unfortunately, these equations do not satisfy the necessary transversality conditions to yield fixed points or solutions, since the second two are obtained by derivation of the first \cite{M39,S42}. However, equations (\ref{ec3}) are linear with respect to $R_J$, and the three values of $R_J$ can be solved as functions of the respective value of the error $\epsilon_i$ for each of the equations:
\begin{equation}
R_{J,i}=\left(\frac{N_i}{E_i \epsilon_i+D_i}\right)^{1/2}, \quad i=1,2,3,
\label{jetradius}
\end{equation}
where
\begin{equation}
D_i=A_i(B_i+C_i {\cal L}), \, i=1,2,3;\quad {\cal L}=\log \left(\frac{\tan \left(\frac{\chi }{2}\right)}{\tan \left(\frac{\alpha}{2}\right)}\right),
\end{equation}
and
\begin{eqnarray}
N_0 = -4 M(\cos (\chi )+1) \left(4 \lambda \left(3 \lambda \cot ^2\left(\frac{\alpha}{2}\right) \cos ^2\left(\frac{\chi }{2}\right)+\right.\right.& \nonumber \\
\left.\left.\left(2 \cos \left(\alpha\right) + 1\right) \sin ^2\left(\frac{\chi }{2}\right)\right)+\left(2 \cos \left(\alpha\right)+1\right) (\cos (\chi )-1)\right);& \nonumber \\
N_1 = \left(16 M \csc \left(\frac{\alpha}{2}\right) \sin (\chi ) \left(4 \lambda \left(3 \lambda \left(\cos \left(\alpha\right)+1\right) (\cos (\chi )+1)+\right.\right.\right.& \nonumber\\
\left.\left.\left. \left(\cos \left(2 \alpha\right)-\cos \left(\alpha\right)\right) \cos (\chi )\right)+2 \left(\cos \left(\alpha\right)-\cos \left(2 \alpha\right)\right) \cos (\chi )\right)\right)/\left(2 \cos \left(\alpha\right)+1\right);&\nonumber \\
N_2 = \left(8 M \csc \left(\frac{\alpha}{2}\right) \left(4 \lambda \left(3 \lambda \left(\cos \left(\alpha\right)+1\right) (\cos (\chi )+\cos (2 \chi ))+\right.\right.\right.& \nonumber \\
\left.\left.\left. \left(\cos \left(2 \alpha\right)-\cos \left(\alpha\right)\right) \cos (2 \chi )\right)+2 \left(\cos \left(\alpha\right)-\cos \left(2 \alpha\right)\right) \cos (2 \chi )\right)\right)/\left(2 \cos \left(\alpha\right)+1\right)&\nonumber;\\
M=\frac{q \cot \left(\frac{\alpha}{2}\right) \left(\left(\lambda-1\right) \cos \left(\alpha\right)+\lambda+1\right)}{\pi  \left(2 \lambda-1\right)};\nonumber\\
A_0=2 \left(\cos \left(\alpha\right)+\cos \left(2 \alpha\right)+1\right) \cos ^2\left(\frac{\chi }{2}\right);&\nonumber\\
B_0=2 \left(\cos \left(\alpha\right)-\cos (\chi )\right) \left(\left(\lambda-1\right) \cos \left(\alpha\right)+\lambda\right);&\nonumber\\
C_0=-8 \cos ^2\left(\frac{\alpha}{2}\right) \sin ^2\left(\frac{\chi }{2}\right) \left(\left(\lambda-1\right) \cos \left(\alpha\right)+\lambda+1\right);&\nonumber
\end{eqnarray}
\begin{eqnarray}
A_1=8 \left(\sin \left(\frac{\alpha}{2}\right)-\sin \left(\frac{3 \alpha}{2}\right)\right) \sin (\chi ),&\nonumber\\
B_1=4 \lambda \cos ^2\left(\frac{\alpha}{2}\right) \left(\cos \left(\alpha\right)-\cos (\chi )\right)+\cos \left(\alpha\right) (2 \cos (\chi )+1)-\cos \left(2 \alpha\right);&\nonumber\\
C_1=2 \cos (\chi ) \left(4 \lambda \cos ^4\left(\frac{\alpha}{2}\right)+\sin ^2\left(\alpha\right)\right);&\nonumber\\
A_2=4 \left(\sin \left(\frac{\alpha}{2}\right)-\sin \left(\frac{3 \alpha}{2}\right)\right.&\nonumber\\
B_2=\lambda \left(2 \left(3 \cos \left(\alpha\right)+\cos \left(2 \alpha\right)+2\right) \cos (\chi )-4 \cos ^2\left(\frac{\alpha}{2}\right) \cos (2 \chi )\right)+&\nonumber\\
2 \cos \left(\alpha\right) \cos (2 \chi )+\left(\cos \left(\alpha\right)-2 \cos \left(2 \alpha\right)+1\right) \cos (\chi );&\nonumber\\
C_2=2 \cos (2 \chi ) \left(4 \lambda \cos ^4\left(\frac{\alpha}{2}\right)+\sin ^2\left(\alpha\right)\right);&\nonumber\\
E_0=\left(2 \cos \left(\frac{\alpha}{2}\right)+\cos \left(\frac{3 \alpha}{2}\right)\right) \csc \left(\frac{\alpha}{2}\right) \left(\left(\lambda-1\right) \cos \left(\alpha\right)+\lambda+1\right);&\nonumber\\
E_1=8 \cos \left(\frac{\alpha}{2}\right) \left(\left(\lambda-1\right) \cos \left(\alpha\right)+\lambda+1\right);\quad E_2=E_1.&
\end{eqnarray}
Setting $\epsilon_i=0$ and defining an error norm as
\beq
\varepsilon(\alpha,\chi;\lambda)= \left((R_{J,1}-R_{J,2})^2+(R_{J,2}-R_{J,3})^2\right)^{1/2}
\eeq
one can investigate the minimum values that $\varepsilon$ can achieve for a given $\lambda$. Figure \ref{f13} illustrates the strong nature of the minimum (a ``creek") that appears for certain relations $\chi=\chi(\alpha;\lambda)$.
The minimum is an extremely well-defined creek that can be located where both $\partial_\alpha \varepsilon=0$ and $\partial_\chi \varepsilon=0$. Figure \ref{f14} gives the locations of the creek on the plane $\{\alpha,\chi\}$ and the values of $\varepsilon_{min}$ for different $\lambda$ values.

Interestingly, the matching angle $\chi$ for the maximum cone angle $\alpha_{max}$ is about $\chi\simeq 1.22$ for all $\lambda$, at least in the range explored. However, the minimum error norm of the jet radius $\varepsilon_{min}$ decreases proportionally to $\lambda^2$. The minimum value of $\varepsilon_{min}$ for a given $\lambda$ is located at a specific cone angle $\alpha^*\cong 0.8\lambda^{1/2}$ with a matching angle $\chi\simeq 2.55$ nearly independent of $\lambda$.

The curves in Figure \ref{f14}(a) can be collapsed defining an abscissa as $x=\alpha \lambda^{-1/2}$, and can be fitted by the following expression:
\beq
f(x)=A x^{\varphi_1}\left(\left(\frac{x}{x_1}\right)^{\delta_1}+1\right)^{\varphi_2/\delta_1}\left(\left(\frac{x}{x_2}\right)^{\delta_2}+1\right)^{-\frac{\varphi_1+\varphi_2}{\delta_2}}
\label{fit}
\eeq
The fitting parameters result
\[
\{A,x_1,x_2,\varphi_1,\varphi_2,\delta_1,\delta_2\} = \{0.71,1.32,1.58,1.05,3.0,13.0,14.0\}.
\]
The closed expression (\ref{fit}) can be used to obtain the approximate matching angle $\chi$ with a good precision as $\chi=\pi-f(\alpha \lambda^{-1/2})$.

\begin{figure}
\centerline{\includegraphics[width=0.70\textwidth]{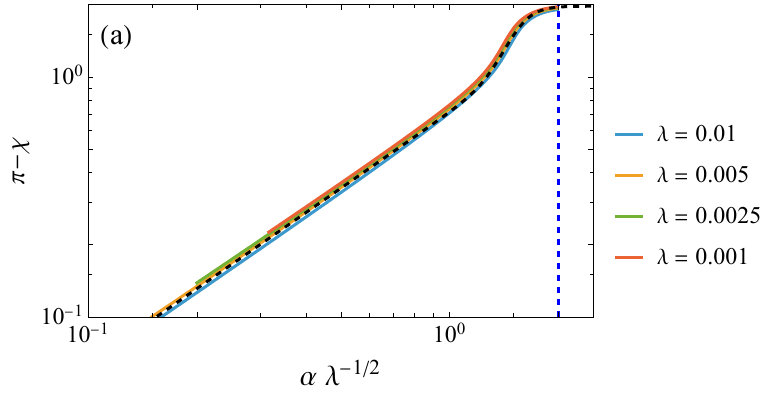}}
\centerline{\includegraphics[width=0.70\textwidth]{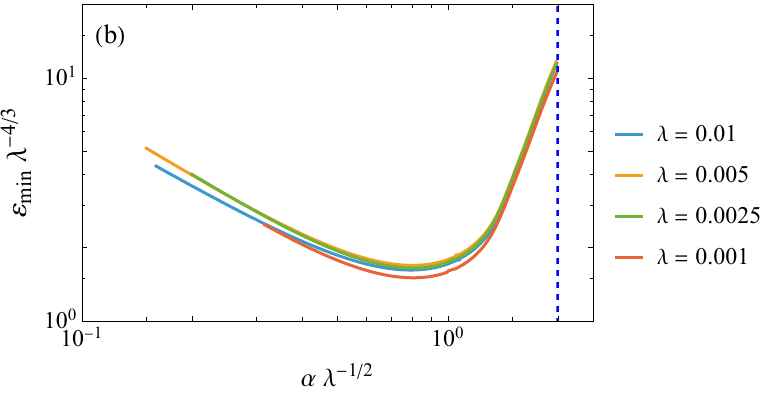}}
  \caption{The collapsed curves of Figure \ref{f14}. (a) The $\pi-\chi$ curves are collapsed by the abscissa $\alpha \lambda^{-1/2}$. The black dashed line is the fitting function $\pi-\chi=f(\alpha \lambda^{-1/2})$. The blue dashed line indicates the maximum cone angle $\alpha_{max}=2 \lambda^{1/2}$. (b) The values of the minima, collapsed by $\varepsilon_{min} \lambda^{-4/3}$ as a function of $\alpha \lambda^{-1/2}$.}
\label{f15}
\end{figure}

In conclusion, the analytical solution produces a local flow solution on the intermediate scale $l$ involving a conical meniscus with angle $\alpha$ related to the strength of the outer flow, defined by a local capillary number $\overline{\rm Ca}$, as shown by the slender body theory. This cone angle is therefore a free parameter for the purposes of this work, but is is actually related to the strength of the macroscopic flow at large scales compared to these where the conical solution here obtained is a valid approximation.

%\vspace*{5mm}
%Acknowledgements...

%\appendix
%\section{}\label{appA}
%This appendix ....

%\bibliography{central}
%\bibliographystyle{plain}

\end{document}